\documentclass[twocolumn,showpacs,preprintnumbers,amsmath,amssymb,showpacs,showcolor]{revtex4}

\usepackage{graphicx}
\usepackage{dcolumn}
\usepackage{bm}
\usepackage{amssymb}
\usepackage{color}
\usepackage{multirow}
\usepackage{makecell}

 \usepackage[colorlinks,linkcolor=blue,citecolor=blue,urlcolor=blue]{hyperref}

 \def\tskip{\setlength{\tskip}{5pt}}
\def\colwidth{\setlength{\colwidth}{3.5in}}

\newcommand{\lsim}{\mathrel{\hbox{\rlap{\lower.55ex\hbox{$\sim$}} \kern-.3em \raise.4ex \hbox{$<$}}}}
\newcommand{\gsim}{\mathrel{\hbox{\rlap{\lower.55ex\hbox{$\sim$}} \kern-.3em \raise.4ex \hbox{$>$}}}}
\newcommand{\beq}{\begin{equation}}
\newcommand{\eeq}{\end{equation}}
\newcommand{\be}{\begin{equation}}
\newcommand{\ee}{\end{equation}}
\newcommand{\bes}{\begin{equation*}}
\newcommand{\ees}{\end{equation*}}
\newcommand{\beqa}{\begin{eqnarray}}
\newcommand{\eeqa}{\end{eqnarray}}
\newcommand{\bea}{\begin{eqnarray}}
\newcommand{\ena}{\end{eqnarray}}

\begin{document}

\title{Detecting Relic Gravitational Waves by Pulsar Timing Arrays: Effects of Cosmic Phase Transitions and Relativistic Free-Streaming Gases}


\author{Xiao-Jin Liu$^{1}$}

\author{Wen Zhao$^{1}$}
\email{wzhao7@ustc.edu.cn}

\author{Yang Zhang$^{1}$}

\author{Zong-Hong Zhu$^{2}$}

\affiliation{ $^1$CAS Key Laboratory for Researches in Galaxies and Cosmology, Department of Astronomy, University of Science and Technology of China, Chinese Academy of Sciences, Hefei, Anhui 230026, China\\
$^2$Department of Astronomy, Beijing Normal University, Beijing 100875, China }

\begin{abstract}

Relic gravitational waves (RGWs) generated in the early Universe form a stochastic GW background, which can be directly probed by measuring the timing residuals of millisecond pulsars. In this paper, we investigate the constraints on the RGWs and on the inflationary parameters by the observations of current and potential future pulsar timing arrays. In particular, we focus on effects of various cosmic phase transitions (e.g. $e^{+}e^{-}$ annihilation, QCD transition and SUSY breaking) and relativistic free-streaming gases (neutrinos and dark fluids) in the general scenario of the early Universe, which have been neglected in the previous works. We find that the phase transitions can significantly damp the RGWs in the sensitive frequency range of pulsar timing arrays, and the upper limits of tensor-to-scalar ratio $r$ increase by a factor $\sim 2$ for both current and future observations. However, the effects of free-steaming neutrinos and dark fluids are all too small to be detected. Meanwhile, we find that, if the effective equation of state $w$ in the early Universe is larger than $1/3$, i.e. deviating from the standard hot big bang scenario, the detection of RGWs by pulsar timing arrays becomes much more promising.

\end{abstract}



\maketitle


\section{Introduction \label{section1}}

Inflation is the most popular scenario of the extremely early Universe \cite{inflation, Early_Universe_1990, Weinberg_2008}. In addition to elegantly solving the flatness puzzles, the horizon puzzles, and the monopole puzzles in the hot big bang universe, inflationary models predict the primordial density fluctuations (scalar perturbations) with the nearly Gaussian distribution, and the nearly scale-invariant power spectrum \cite{scalar}, which have been strongly supported by the recent observations on the anisotropies of cosmic microwave background (CMB) radiation \cite{wmap-results,planck-results,Planck_parameters_2015}, and the distributions of the galaxies in various large-scale structure observations.

In the inflation scenario, a stochastic background of relic (primordial) gravitational waves (RGWs) in full  frequency range was inevitable produced due to the superadiabatic amplification of zero-point quantum fluctuations of the gravitational field \cite{grishchuk}, which provides the unique window to study the physics in the early Universe before the big bang nucleosynthesis (BBN) stage. Nowadays, the detection of RGWs in different frequencies has been carried out by different methods. In the lowest frequency range with $f<10^{-15}$Hz, it can be detected by their imprints in the CMB anisotropies in temperature and polarizations \cite{cmb-method}. The recent {Planck} observations on the CMB temperature and E-mode polarizations give the tightest constraint on the tensor-to-scalar ratio $r<0.11$ \cite{Planck_parameters_2015}, which is consistent with the result $r<0.12$ obtained from the joint analysis of BICEP2/{Keck Array} and {Planck} B-mode polarization data \cite{Bicep2_planck_2015}. In the near future, one anticipates that the detection limit can arrive at $r=0.01$ for various ground-based or balloon-borne experiments (including Keck/BICEP3, POLARBEAR, SPT-3G, ACTPOL, CLASS, QUBIC, QUIJOTE, QUIET, Simons Array, EBEX, PIPER, SPIDER et al.). While for the space-based missions, such as CMBPOL, LiteBird, COrE, PRISM, PIXIE, the detection limit can be $r=0.001$ \cite{cmb-mission}.

In the high frequency range, i.e. $f\in(10^{-4}, 10^{4})$Hz, RGWs are detected by the ground-based and space-based laser interferometer gravitational-wave observatories. So far the most stringent bound is obtained by the joint analysis of the 2009-2010 LIGO and Virgo data, which gives energy density of RGWs $\Omega_{\rm gw}(f)<5.6\times10^{-6}$ at frequency band spanning 41.5-1726Hz \cite{LIGO_correlation}. In the future, the detection limits at $f\sim 100$Hz are expected to be $\sim 10^{-9}$ for the second-generation gravitational-wave detectors, such as AdvLIGO, AdvVirgo, KAGRA and so on \cite{aligo}. For the third-generation detectors, such as Einstein Telescope, this limit could arrive at $\sim 10^{-11}$ \cite{et}. The space-based detectors are sensitive to the lower-frequency gravitational wave (about 0.1 mHz to 1 Hz). For the future eLISA/NGO mission, the detection ability is hopeful to be $\sim 4\times 10^{-10}$ at $f=4$ mHz \cite{lisa}. In addition, the limit can be even improved by 5-10 orders by BBO \cite{bbo}, DECIGO \cite{decigo} and ASTROD \cite{astrod} in the far future.

In this paper, we shall focus on the RGWs in the median frequency range. In the frequency band $f\in(10^{-9},10^{-7})$Hz, the gravitational waves are probed through the pulsar timing arrays (PTAs). It is well known that, the millisecond pulsars are the very stable clocks. The tiny timing residuals of these pulsars are caused by some intrinsic or environmental noises, as well as gravitational waves \cite{detweiler}. It was realized that the timing residuals from an array of pulsars could be analyzed coherently to separate GW-induced residuals from other effects \cite{Hellings_downs_1983, Foster_1990}. So, it provides the excellent way to detect RGWs in the median frequency range. Nowadays, these observations are carried out by three groups (PPTA, EPTA, NANOGrav). Recently, all teams released their latest timing results \cite{Parkes_2013,EPTA_2015,NANO_2015}, and the tightest constraint is obtained by the PPTA team, they placed a $95\%$ upper limit on the strain amplitude (at a frequency of yr$^{-1}$) in the power-law model of $A_{\rm gw}<1.0\times10^{-15}$ for spectral index $-2/3$ \cite{Parkes_2013}. These bounds could be significantly improved by the potential observations of future International Pulsar Timing Array (IPTA), Five-hundred-meter Aperture Spherical Radio Telescope (FAST) in China and the planned Square Kilometer Array (SKA) projects \cite{Zhao_et_al_2013}. These upper limits of GW background have also been applied to constrain the RGWs and on the inflation physics \cite{EPTA_2015,NANO_2015,Zhao_2011,Zhao_et_al_2013,tong2014,tong2009}. In all these previous works, a simple evolution model is used to calculate the RGW power spectrum, where the effects of free-streaming fluids and various cosmic phase transitions are neglected. However, it was known that, the relativistic free-streaming gas (such as the neutrinos in the Universe) gives rise to an anisotropic term in the evolution equation of gravitational waves, which can significantly damp the RGW spectra at $f>10^{-16}$Hz \cite{Weinberg_2005,Zhao_2009}.
Another effect is caused by the successive changes in the relativistic degrees of freedom during the radiation-dominant stage, i.e. the QCD transition, $e^+e^-$ annihilation, the electroweak phase transition and so on. During these transitions, the evolution of scale factor was altered compared with the standard radiation-dominant stage, which left the imprints in the RGW spectrum at $f>10^{-10}$Hz \cite{Watanabe_komatsu_2006,Zhao_2007,wang2008}. So, both effects can change the RGWs at the median frequency range. In this paper, we shall investigate in details the effects of free-steaming gas and cosmic phase transitions on the RGW spectrum in a general cosmological scenario. In particular, we shall focus on their influences on the detection of RGWs by the current and future PTAs and the corresponding cosmological implications.

The outline of this paper is as follows. In Sec. II, we introduce the RGWs in the accelerating Universe by considering the effects of free-steaming gas, cosmic phase transitions and unusual equation of state. In Sec. III and Sec. IV, we discuss the constraints on the inflationary parameters by the current and future PTA observations. Section V summarizes the main results of this paper.


\section{Relic gravitational waves \label{section2}}

The action of gravitational wave $h_{ij}$ is \cite{Boyle_steinhardt_2008}
\begin{equation}
 \label{action_of_RGW}
  S=\int d\tau d^{3}x\sqrt{-\bar{g}}\Big[\frac{-\bar{g}^{\mu\nu}}{64\pi G}\partial_{\mu}h_{ij}\partial_{\nu}h^{ij}+\frac{1}{2}\Pi_{ij}h^{ij}\Big],
  \end{equation}
where $\bar{g}_{\mu\nu}$=$\text{diag}\{ -a^{2}, a^{2}, a^{2}, a^{2}\}$ is Friedmann-Lema\^itre-Robertson-Walker metric, with $a$ the scale factor, $G$ the Newtonian gravitational constant and $\tau$ the conformal time. The term $\Pi_{ij}$ is anisotropic stress, which includes the contribution of large-scale magnetic field \cite{Maartens_tsagas_ungarelli_2001}, free-streaming relativistic particles (e.g. neutrinos after their decoupling) \cite{Weinberg_2005, Dicus_Reoko_2005} and so on.

By applying Euler-Lagrange equation to (\ref{action_of_RGW}), we can obtain the equation of motion for $h_{ij}(\vec{x}, \tau)$. Decompose the perturbation $h_{ij}(\vec{x}, \tau)$ and $\Pi_{ij}(\vec{x}, \tau)$ by Fourier transformation, then we  obtain the equation of evolution for the mode with conformal wavenumber $k$:
\begin{equation}
\label{equation_of_evolution}
     h_{k}^{''}(\tau) + 2\frac{a'}{a}h_{k}^{'}(\tau)+k^{2}h_{k}(\tau)=16\pi G a^{2}(\tau) \Pi_{k}
(\tau),
\end{equation}
where the prime ($'$) denotes derivative with respect to $\tau$ and $\Pi_{k}$ is the Fourier component of $\Pi_{ij}(\vec{x}, \tau)$.

The power spectrum of RGWs at $\tau$ is defined by \cite{Boyle_steinhardt_2008}
\begin{equation}
  \label{power_spectrum_from_hk}
   P_{h}(k, \tau)=64\pi G \frac{k^{3}}{2\pi^{2}}\big |h_{k}(\tau)\big|^{2},
  \end{equation}
while the initial power spectrum (spectrum at $\tau=0$) is usually parameterized by a power-law form \cite{Peiris_komatsu_et_al_2003}
\begin{equation}
 \label{power_spectrum_of_power_law}
   P_{h}(k, 0)=r A_{R}(k_{0})\Big(\frac{k}{k_{0}}\Big)^{n_{t}},
   \end{equation}
where $r$ is tensor-to-scalar ratio, and $A_{R}(k_{0})$ is the value of scalar power spectrum at $k=k_{0}$, with $k_{0}$ a pivot number, and $n_{t}$ the power index of tensorial  spectrum. Throughout this paper, we use $k_{0}=0.002$ Mpc$^{-1}$, and $A_{R}(k_{0})= 2.371\times10^{-9}$ (this result is obtained by converting $A_{R}=2.139\times10^{-9}$ \cite{Planck_parameters_2015} at $k=0.05$ Mpc$^{-1}$ into that at $k=0.002$ Mpc$^{-1}$).

The tensor-to-scalar ratio $r$ is currently constrained to be $r<0.11$ (95\%.C.L.)\cite{Planck_parameters_2015}. For de-Sitter inflation, the power spectrum becomes flat when the modes leave the horizon, so $n_{t}=0$ \cite{Dodelson_2003}. For slow-roll inflation, due to the null energy condition, $n_{t}<0$ \cite{Baumann_McAllister_2014}, but non-canonical inflation or the extension of slow-roll model may predict $n_{t}>0$ \cite{Cai_yifu_2014}.

It is convenient to convert power spectrum into dimensionless spectrum of energy density \cite{Boyle_steinhardt_2008}
\begin{equation}
 \label{definition_of_spectrum_of_energy_density}
  \Omega_{\text{gw}}(k, \tau)=\frac{1}{\rho_{\text{c}}}\frac{d\rho_{\text{gw}}}{d\ln k}=\frac{k^{2}P_{h}(k, \tau)}{12a^{2}(\tau)H^{2}(\tau)},
  \end{equation}
where $\rho_{\text{gw}}$ is the energy density of gravitational waves, and $\rho_{\text{c}}=3H^{2}/\big(8\pi G\big)$ is the critical energy density of the Universe (we set $c=1$). This expression is equivalent to Eq. (7) in our previous work
\cite{Zhao_et_al_2013}.

Throughout this paper, we use the standard model of cosmology. The cosmological parameters adopted in this paper are listed here: $\Omega_{\text{m}}=0.308$, $z_{\rm eq}=3365$, current conformal time $\tau_{0}=1.41\times10^{4}$ Mpc, present scale factor $a(\tau_{0})=1$, current Hubble parameter
$H(\tau_{0})=67.8 $km s$^{-1}$ Mpc$^{-1}$, and the reduced Hubble parameter $h=H/\big(100 \text{km s}^{-1}\text{Mpc}^{-1}\big)$ \cite{Planck_parameters_2015}.

Because the equation of evolution (\ref{equation_of_evolution}) is difficult to be solved analytically (see \cite{Zhang_et_al_2006}, for example), it is useful to introduce transfer function $T(k, \tau)$ as follows \cite{Turner_et_al_1993}
\begin{equation}
  \label{definition_of_transfer_function}
    T(k,\tau)=\frac{h_{k}(\tau)}{h_{k}(0)},
    \end{equation}
where $h_{k}(0)$ is the initial amplitude of the mode with wavenumber $k$. Now all the effects of evolution in the post-inflation stage are attributed to the transfer function and we split the transfer function into several parts to account for different damping factors \cite{Boyle_steinhardt_2008, Zhao_zhang_2006}
\begin{equation}
  \label{separation_of_transfer_function}
   T(k, \tau)=T_{z}(k, \tau)\times T_{\text{PT}}(k, \tau)\times T_{\nu}(k, \tau)\times T_{\text{EoS}}(k, \tau),
   \end{equation}
where $T_{z}$ represents the transfer function of the cosmic expansion (the reason of cosmological redshift $z$), while $T_{\text{PT}}$ is for cosmic  phase transitions in hot Universe, $T_{\nu}$ is for free-streaming particles(e.g. neutrinos $\nu$) and $T_{\text{EoS}}$ is for unusual equation of state (EoS) before BBN epoch.

Inserting (\ref{power_spectrum_from_hk}), (\ref{power_spectrum_of_power_law}) and (\ref{definition_of_transfer_function}) into (\ref{definition_of_spectrum_of_energy_density}), the present spectrum of energy density becomes
\begin{equation}
 \label{expanded_spectrum_of_energy_density}
   \Omega_{\text{gw}}(k, \tau_{0})=\frac{r k^{2}A_{R}(k_{0})}{12H^{2}(\tau_{0})}\Big(\frac{k}{k_{0}}\Big)^{n_{t}}\times T^{2}(k, \tau_0).
\end{equation}
From (\ref{separation_of_transfer_function}) and (\ref{expanded_spectrum_of_energy_density}), we could see that the transfer functions determine the strength of RGW spectrum to some extent and that they are closely related to the physics after inflation. Now we would like to discuss the transfer functions one by one.

\subsection{RGWs in the accelerating Universe}

RGWs were stretched out of the horizon during the inflationary epoch and re-entered the horizon in the following epochs of decelerating expansion and recent accelerating expansion \cite{Dodelson_2003,Weinberg_2008}.

When stretched out of the horizon, RGWs started to freeze out and the amplitude kept constant, because out of the horizon there are no anisotropic tensor to damp the RGWs even when there are some particles whose free path may be comparable to the horizon \cite{Weinberg_2005,Weinberg_2008}. When the Universe exited inflation, the horizon started to increase and the modes with high $k$ (thus short wavelength) started to re-enter the horizon. The criterion of horizon crossing is $a(\tau)H(\tau)=k$ \cite{Stewart_Lyth_1993}. After re-entering the horizon, RGWs are mainly damped by the expansion of the Universe. According to (\ref{equation_of_evolution}), when the anisotropic term $\Pi_{ij}$ is neglected, the amplitude of modes in the inner part of the horizon ($a(\tau) H(\tau)\ll k$) would satisfy
\begin{equation}
 \label{h_inverse_scale_factor}
   h_{k}( \tau)\approx \frac{h_{k}(0)}{a(\tau)}.
   \end{equation}
So the evolution of RGWs is closely related to the history of expansion. Turner et al  \cite{Turner_et_al_1993} formulated the transfer function in the decelerating expansion by fitting it to the numerically integrated results, while Zhang et al noticed the effect of accelerating expansion \cite{Zhang_et_al_2005, Kuroyanagi_et_al_2010}. Now the transfer function in the expanding Universe is the product of $T_{\text{RM}}$ (radiation and matter dominated era) and $T_{\Lambda}$ ($\Lambda$ dominated era), or precisely \cite{Turner_et_al_1993, Zhang_et_al_2005, efstathiou2006, Giovannini2010, Kuroyanagi_et_al_2010}
\begin{equation}
  \label{transfer_function_redshift}
  T_{z}(k, \tau_{0})=\frac{3 \Omega_{\text{m}}j_1(k\tau_0)}{k\tau_{0}}\sqrt{1.0+1.36\frac{k}{k_{\text{eq}}}+2.50\big(\frac{k}{k_{\text{eq}}})^{2}},
  \end{equation}
where $k_{\text{eq}}=0.073 \Omega_{\text{m}} h^{2}$ Mpc$^{-1}$ is the wavenumber corresponding to the mode that entered the horizon at the equality of matter and radiation,  and $j_1(k\tau_0)$ is the spherical Bessel function of the first kind.

\begin{figure*}[!htb]
 \begin{center}
  \includegraphics[width=12.4cm]{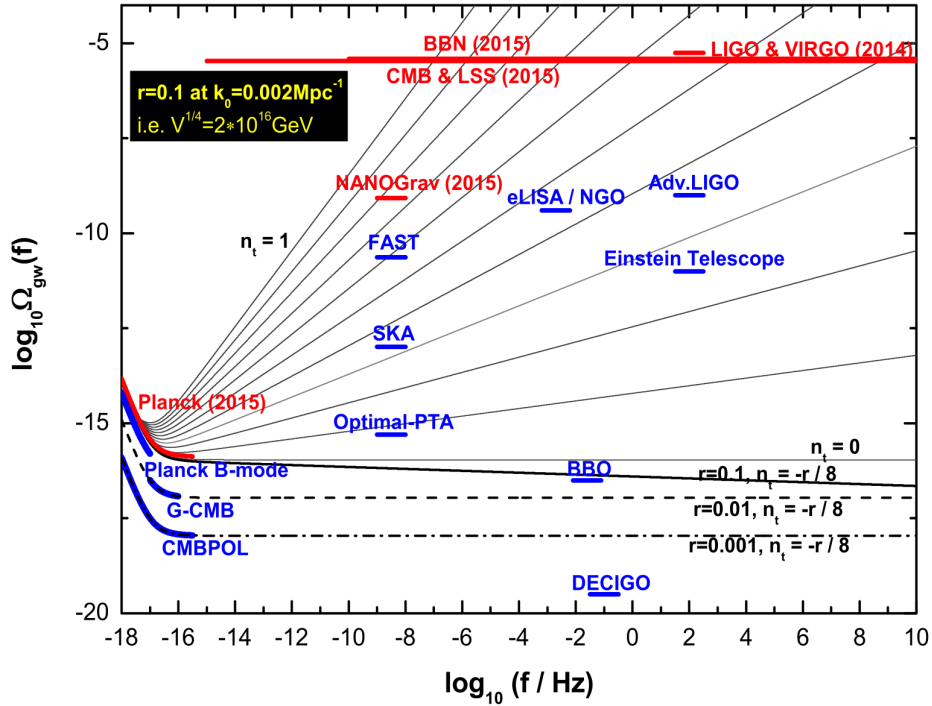}
\end{center}
 \caption{\small The energy density spectrum of RGWs $\Omega_{\rm gw} (f)$ for different inflationary models, where we only consider the damping effect caused by the cosmic expansion in the standard hot big-bang Universe. In this figure, we also show the current constraints on RGWs by different observations (red lines) and the potential constraints by future observations (blue lines). The details of these constraints can be found in the main text.}
 \label{figure1}
 \end{figure*}

In Fig.\ \ref{figure1}, we plot the energy density spectrum of RGWs by only considering the damping effect of the cosmic expansion, where different spectral indices $n_t$ are considered. For the comparison, we have also plotted the current and potential constraints/detection of RGWs by various observations.

\subsection{Effects of cosmic phase transitions: $e^{+}e^{-}$ annihilation, QCD transition and SUSY breaking}

When the hot Universe gradually cooled down in the radiation-dominant era, massive particles became less relativistic and contributed less to the energy density of radiation. Typical cases are  the breaking of supersymmetry (SUSY) and the combination of quarks into hadrons (phase transition of quantum chromodynamics i.e. QCD phase transition). Besides that, the annihilation of electrons and positrons also induced big change in the energy density of radiation. Here, we generally call all these three effects and similar effects that could induce the change of energy density of radiation cosmic phase transition.

The cosmic phase transitions change the energy density of radiation thus the Hubble parameter $H(\tau)=a'/a^2$, and affect the amplitude of RGWs through Eq. (\ref{h_inverse_scale_factor}). The transfer function of phase transitions is \cite{Zhao_2007, Watanabe_komatsu_2006}

\begin{equation}
  \label{transfer_function_of_phase_transition}
  T_{\text{PT}}(k, \tau_{0})=\Big(\frac{g_{\star}(T_{k})}{g_{\star 0}}\Big)^{\frac{1}{2}}\Big(\frac{g_{\star s}(T_{k})}{g_{\star s0}}\Big)^{-\frac{2}{3}},
   \end{equation}
where $g_{\star}(T_{k})$ is effective degree of freedom (here effective degree of freedom is obtained by converting all particles into effective photons, see Chapter 3.3 in \cite{Early_Universe_1990}) for total energy density and  $g_{\star s}(T_{k})$ is the effective degree of freedom for entropy density at temperature $T_{k}$, while $g_{\star 0}$ and $g_{\star s0}$ are the present values. The parameter $T_{k}$ is the temperature at which the mode $k$ re-enters the horizon.

By use of the condition of horizon crossing $aH=k$ and the conversation of entropy $g_{\star s}a^{3}T^{3}=$ constant, we can convert temperature into wavenumber:
\begin{equation}
 \begin{split}
   \label{transformation_from_t_to_k}
   k&=\frac{H(\tau_{0})T(\tau_{0})}{T_{k}}\Big(\frac{g_{\star s0}}{g_{\star s}}\Big)^{\frac{1}{3}}\\
   &\times\sqrt{\Omega_{\text{m}}\frac{g_{\star s}}{g_{\star s0}}\Big(\frac{T_k}{T(\tau_{0})}\Big)^3+\frac{\Omega_{\text{m}}}{1+z_{\text{eq}}}\frac{g_{\star}}{g_{\star 0}}\Big(\frac{T_k}{T(\tau_0)}\Big)^4+\Omega_{\Lambda}},
   \end{split}
  \end{equation}
where $T(\tau_{0})=2.7255$ K is the current CMB temperature \cite{Fixsen_2009} and $z_{\text{eq}}$ is the redshift at $\Omega_{\text{m}}(\tau_{\rm eq})=\Omega_{\text{r}}(\tau_{\rm eq})$. We can further convert wavenumber into frequency: $k=2\pi f$.

The transfer function of phase transitions depends on not only the temperature but also the physical characteristics of particles, such as mass, degree of freedom and the statistical features (bosons or fermions) and so on. We follow the list of particles in \cite{Watanabe_komatsu_2006} but update the data according to \cite{pdg_oliver_2014}. The temperature when particles decouple from the others is also very important. Here we adopt instantaneous decoupling of neutrinos from the radiation at $T=1.5$ MeV, and that electrons annihilate with positrons immediately at $T=0.1$ MeV \cite{Mukhanov_2005}. The phase transition of QCD is treated as first order transition at 155 MeV \cite{Hot_QCD_2014}.

\begin{figure}[!htb]
  \begin{center}
    \includegraphics [width=8.5cm]{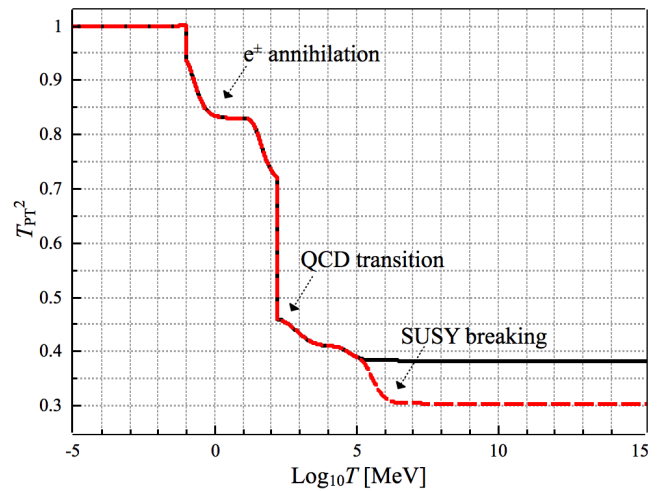}
   \end{center}
 \caption {\small The dependence of $T_{\text{PT}}^{2}(k, \tau_{0})$ on temperature: The solid line (black) is the case without SUSY particles, while the dashed one (red) is the case containing them. On the dashed line, from right to left there are three typical steps, which are the evidence of breaking of SUSY ($\sim10^{6}$ MeV), phase transition of quarks ($\sim$ 155 MeV) and the annihilation of $e^{+} e^{-}$($\sim$ 0.1 MeV). The figure is plotted without considering the dark fluid (introduced in section \ref{introduction_of_dark_fluid}). }
 \label{phase_transition_transfer_function1}
 \end{figure}

Fig.\ \ref{phase_transition_transfer_function1} shows the square of transfer function $T_{\text{PT}}^{2}(k, \tau_{0})$ at different temperature.  When the cosmic phase transitions are considered, the energy density spectrum of RGWs can be reduced by about $60\%$ at $10^{2}\sim10^{5}$ MeV and by about $20\%$ at $0.1\sim 10$ MeV. When we include the SUSY particles, the spectrum could be damped as much as $70\%$ at $T>10^{6}$ MeV. However, SUSY breaking is beyond the detecting ability of PTAs ($52\sim3.9\times10^3$ MeV). So we will not consider SUSY particles later.

\subsection{Effects of relativistic free-streaming gases: neutrinos and dark fluids}

\subsubsection{Neutrinos}

Now, let us discuss the anisotropic term in (\ref{equation_of_evolution}).  As the cosmic magnetic filed is very small, we will ignore the magnetic sources \cite{Maartens_tsagas_ungarelli_2001}. When neutrinos decouple from the rest of radiation, the free streaming of neutrinos contributes to the anisotropic stress $\Pi_k(\tau)$, which plays the role of friction in the equation of motion and damps the amplitude of RGWs, so it is necessary to study the damping effect of neutrinos.

In very high precision, the transfer function of relativistic free-streaming particles is \cite{Boyle_steinhardt_2008, Dicus_Reoko_2005}
\begin{widetext}
\begin{equation}
 \label{damping_factor_of_neutrinos}
  T_{\nu}(k, \tau_{0})=\frac{15(324 135 000- 48 118 000 f_{\nu}+3 152 975 f_{\nu}^{2}-55 770 f_{\nu}^{3}+14 406f_{\nu}^{4} )}{343(15+4f_{\nu})(50+4f_{\nu})(105+4f_{\nu})(180+4f_{\nu})},
  \end{equation}
\end{widetext}
where $f_{\nu}$ is the fraction of free-streaming particles' energy density over the total energy density. This formula applies for modes satisfying $\sqrt{2}k\gg k_{\text{eq}}$. When $k$ decreases and becomes comparable with $k_{\text{eq}}$, this formula does not apply any more. However, as $k$ decreases, the transfer function will gradually increases to 1 \cite{Stefanek_repko_2013}.

Because the sensitive band of PTAs is $10^{-9}\sim10^{-7}$ Hz, which is much higher than $k_{\text{eq}}$, we would not consider the behavior of free-streaming neutrinos near $k_{\text{eq}}$, thus (\ref{damping_factor_of_neutrinos}) is enough for us and we take it  as the transfer function of free-streaming particles.

When there are neutrinos only, $f_{\nu}$ is the fraction of energy density of free-streaming neutrinos. Before their decoupling from the radiation, $f_{\nu}=0$, thus $T_{\nu}(k, \tau_{0})=1$, so neutrinos could not damp the RGWs. After the decoupling, $f_{\nu}$ started to evolve with the Universe and depends on the temperature.

In Fig.\ \ref{fraction_neutrinos_and_dark_fluid}, we plot the fraction of neutrinos' energy density. We find that $f_{\nu}$ can be roughly divided into two constant stages $f_{\nu}=0.491$ with $T^2_{\nu}(k, \tau_0)=0.589$ (before e$^{+}e^{-}$ annihilation) and $f_{\nu}=0.405$ with $T^2_{\nu}(k, \tau_0)=0.645$ (after e$^{+}e^{-}$ annihilation). Therefore, our analysis is consistent with that in \cite{Weinberg_2005, Boyle_steinhardt_2008}

It should be noted that the dip at $\log_{10} (T/\text{MeV}) \in (-1.2, -1.0)$ and the spike at $\log_{10}(T/\text{MeV})\in(-1.0, -0.6)$ are artificial signatures. They are due to the imperfect deal with annihilation (we assume instantaneous annihilation here, which is not the real situation). The same artifacts appears in \cite{Watanabe_komatsu_2006}, while \cite{Kuroyanagi_chiba_2009} explains this in its Appendix C.

Fig.\ \ref{three_transfer_function} plots the transfer function of cosmological redshift, neutrinos and cosmic phase transitions. Apparently, $T_z$ dominates all the damping effects in a wide range from $10^{-15}$Hz to $10^4$Hz, while neutrino streaming would decorate the spectrum in the low frequency band ($<10^{-10}$Hz) and the cosmic phase transitions modifies RGWs in the high frequency range ($>10^{-12}$Hz). As we could see, phase transitions show their importance especially around $10^{-8}$ Hz, which well resides in the sensitive band of PTAs.

\subsubsection{Dark fluids}
\label{introduction_of_dark_fluid}

In the Standard Model of particle physics, there are three species of neutrinos. But when we consider the non-instantaneous decoupling of neutrinos, the effective number of neutrino species $N_{\text{eff}}=3.046$ \cite{Mangano_2002}. The current constraint from CMB and matter power spectra is $N_{\text{eff}} < 3.376$ (95\% confidence) \cite{Pagano_2015}, while the BBN observations give $N_{\text{eff}} < 3.41$ (95\% confidence) \cite{Cyburt_2015}, so tension is still possible between theory and observation. Weinberg \cite{Weinberg_2013} proposed massless bosons to relax the tension and found $\Delta N_{\text{eff}}=0.39$ for bosons decouple at $T > 100$MeV.  Thus the current CMB observation can not exclude the existence of extra neutrino-like particles.

\begin{figure} [!htb]
 \begin{center}
   \includegraphics [width=8.5cm]{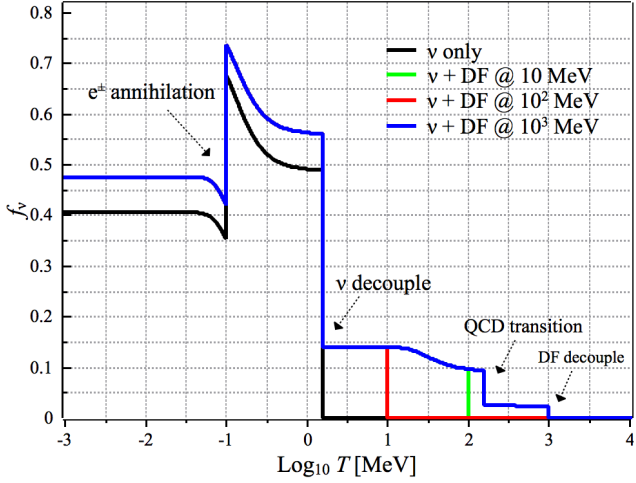}
   \end{center}
 \caption{\small The fraction $f_{\nu}$ of free-streamings' energy density over the total energy density: The black curve is the case without dark fluid, while the red, green and blue ones contain dark fluids which decouple at 10, $10^{2}$ and $10^{3}$ MeV respectively. Neutrinos start to decouple at 1.5MeV and $e^{+}e^{-}$ annihilate at 0.1 MeV, while phase transition of QCD happens at 155 MeV. When there are neutrinos only, the annihilation at $T=0.1$ MeV splits $f_{\nu}$ into two stages $f_{\nu}\approx 0.491$ and $0.405$. When dark fluids are included, $f_{\nu}$ increases and has more stages: $f_{\nu }\approx 0.13$ for $T\in (1.5, 155$) MeV and $f_{\nu}\approx 0.025$ for $T\in (155, 1000)$ MeV.}
  \label{fraction_neutrinos_and_dark_fluid}
\end{figure}

\begin{figure}[!htb]
 \begin{center}
  \includegraphics [width=8.5cm]{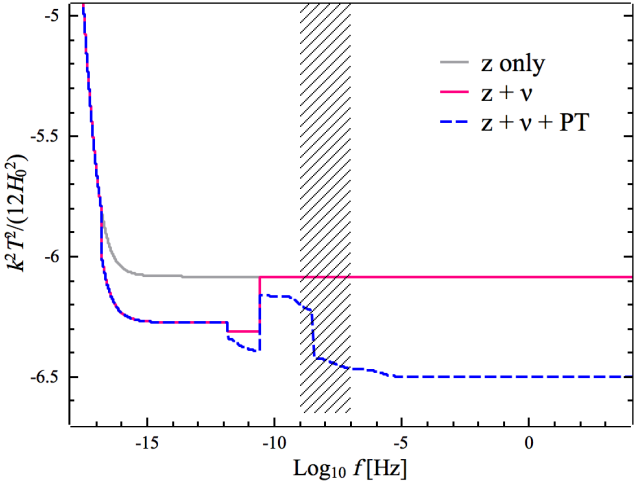}
   \end{center}
  \caption{\small The compilation of transfer functions from $T_z(k, \tau_0)$, $T_{\nu}(k, \tau_0)$ and $T_{\text{PT}}(k, \tau_0)$. The grey line is for $T_z$, while the red one is for the product of $T_z$ and $T_\nu$ and the dotted blue one combines all the three damping factors. The shaded area denotes the sensitive band of pulsar timing arrays. No SUSY particle is considered here.}
  \label{three_transfer_function}
\end{figure}

Here, we assume the existence of extra fermions to address the $\Delta N_{\text{eff}}$ problem. Since massive particles freeze out at low temperature and do not contribute to $N_{\text{eff}}$, we would like to assume the particles are massless. Due to the constraints on $N_{\text{eff}}$, it is well justified to set the physical degree of freedom of dark fluid to be two (one for particle, the other for anti-particle), then $\Delta N_{\rm eff}=1$. To consider the possibilities of different particles, we shall consider different decoupling temperature. Since the PTA band $f\in(10^{-9}, 10^{-7})$ Hz corresponds to $52\sim 3.9\times10^{3}$ MeV,  we consider three decoupling temperature $10, 10^2$ and $10^3$ MeV in our discussion. As the extra particles can be any possible particles beyond current observations, we simply call them dark fluids (DFs).

Fig.\ \ref{fraction_neutrinos_and_dark_fluid} shows the fraction of free-streaming particles including dark fluid. When we include DF, $f_{\nu}$ becomes the fraction of all free-streaming particles' energy density. Both neutrinos and dark fluid contribute to $f_{\nu}$. Rich features occur in the fraction $f_{\nu}$ of free-streaming particles. In Fig.\ \ref{fraction_neutrinos_and_dark_fluid}, we consider three different decoupling temperature ($T=10, 10^{2}$ and $10^{3}$ MeV) of dark fluid. Besides the two stages $f_{\nu}\approx 0.476$ with  $T^2_{\nu}(k, \tau_0)=0.599$ and $f_{\nu}\approx 0.563$ with $T^2_{\nu}(k, \tau_0)=0.547$ separated by annihilation of $e^{+}e^{-}$ at $T=0.1$ MeV, one more stage ($f_{\nu}\approx 0.13, T_{\nu}^2(k, \tau_0)=0.867$) appears when dark fluid decouples at $10$ or $10^{2}$ MeV, and a third stage ($f_{\nu}\approx 0.025$ with $T_{\nu}^2(k, \tau_0)=0.973$) occurs when dark fluid decouples at $10^{3}$ MeV. Thus dark fluid would damp the RGWs at higher frequency.


\begin{figure}
 \begin{center}
  \includegraphics[width=8.5cm]{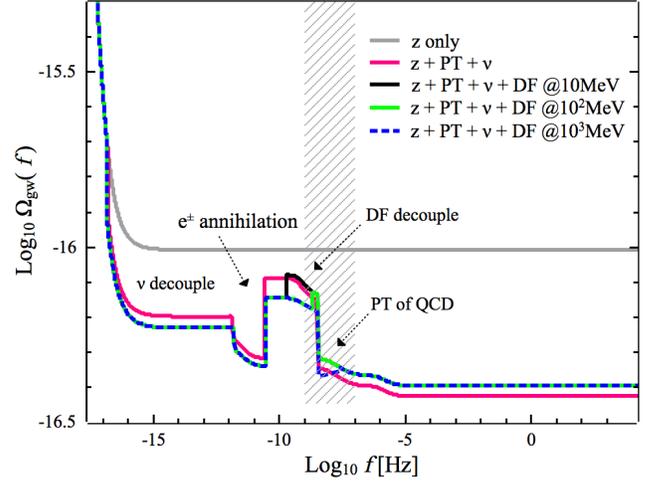}
  \end{center}
  \caption{\small The energy density spectrum of RGW including $T_{z}(k, \tau_{0}),  T_{\text{PT}}(k, \tau_{0})$ and $T_{\nu}(k, \tau)$. In this figure, $r$ is chosen to be $0.1$ and $n_{t}=0$. The top grey line is the spectrum considering the effect of cosmological redshift $T_{z}(k, \tau_{0})$ only, while the pink curve includes free-streaming neutrinos $T_{\nu}(k, \tau_{0})$ and cosmic phase transitions $T_{\text{PT}}(k, \tau_{0})$. Besides damping from $T_{z}(k, \tau_{0})$, $T_{\rm PT}(k, \tau_0)$ and $T_\nu(k, \tau_0)$, the black, green and the dashed blue line also include the effects of dark fluid, which is assumed to decouple at $10, 10^{2}$ and $10^{3}$MeV respectively. The shaded area denotes the sensitive band $10^{-9}\sim10^{-7}$Hz of PTAs. (In this plot, we do not consider SUSY particles.)}
   \label{spectrum_of_energy_density_all_effects}
  \end{figure}

By use of the expression of spectrum (\ref{expanded_spectrum_of_energy_density}), the transfer function (\ref{transfer_function_redshift}), (\ref{transfer_function_of_phase_transition}) and (\ref{damping_factor_of_neutrinos}), we can plot the energy density spectrum of RGWs for given $r$ and $n_{t}$.

Fig.\ \ref{spectrum_of_energy_density_all_effects} shows the spectrum of energy density for $r=0.1$, $n_{t}=0$. In this plot we could see the damping effects caused by cosmological redshift, the cosmic phase transitions and the free-streaming neutrinos/DF.  Neutrinos damp the spectrum at $10^{-16}<f< 10^{-10}$Hz, beyond the sensitive band of PTAs, thus PTAs cannot detect any relic signals from neutrinos. The cosmic phase transitions mainly damp the spectrum at $f> 10^{-9}$ Hz by amount of $60\%$. The big jump at $f\sim10^{-8}$ Hz is the signature of QCD phase transition ($T=155$ MeV).

In Fig.\ \ref{spectrum_of_energy_density_all_effects}, we find that DFs have three kinds of signature in the spectrum. Firstly, steps occur after their decoupling from the radiation. Secondly, the spectrum at $f<10^{-10}$ Hz is damped more. Lastly, spectrum at $f>10^{-8}$ Hz is raised.

At the moment of decoupling ($T=10, 10^{2}, 10^{3}$ MeV), big jumps occur, as the free-streaming dark fluids start to damp the spectrum. In the range of the decoupling of neutrinos ($10^{-16}\sim 10^{-10}$ Hz), the appearance of  dark fluid increases the fraction of free-streaming particles thus damps the spectrum more. This result is consistent with our intuition. But at higher frequencies ($f>10^{-8}$Hz) dark fluids raise the spectrum when compared to the case without dark fluids.

The reason why dark fluids raise the spectrum at high frequencies resides in the massless feature of dark fluid. We can clearly see this from the transfer function of phase transition (\ref{transfer_function_of_phase_transition}). Massless particles can  keep relativistic at low temperature and always contribute to the effective degree of freedom for energy and entropy density. So the addition of dark fluids increases the value of $g_{\star s0}$ and $g_{\star s}(T_{k})$. As $g_{\star}(T_{k})\approx g_{\star s}(T_{k})$ for all $T_{k}$ (see Chapter 3.3 in \cite{Early_Universe_1990}), then
\begin{equation}
 \label{dark_fluid_raise_spectrum}
  T_{\text{PT}}(k, \tau_{0})\approx \Big( \frac{g_{\star s}(T_{k})}{g_{\star s0}} \Big)^{-\frac{1}{6}}.
   \end{equation}
At high temperature (which corresponds to high frequency) $g_{\star s}(T_{k})\gg g_{\star s0}$, so the increase of $g_{\star s0}$ dominates and makes $T_{\text{PT}}$ increase a bit.

\subsection{Effects of general equation of state}

After inflation, the Universe should experience a period of reheating to create the matter and dark matter we see today. Some time later, dark matter should decouple from the rest of particles \cite{Weinberg_2008}. It is suspected that this epoch before BBN can be dominated by massive particles and cause deviation from ideal fluid \cite{Boyle_steinhardt_2008, Boyle_buonanno_2008}, thus unusual EoS may occur. EoS could deeply change the shape of spectrum of energy density through scale factor as we would see below.

The relation between scale factor and conformal time could be generally parameterized by
\begin{equation}
  a\propto \tau^{1+\beta},
  \end{equation}
where $\beta$ is related to the EoS parameter $w$ through
\begin{equation}
   \beta=\frac{1-3w}{1+3w},\qquad \text{with}\quad w=\frac{p}{\rho}.
    \end{equation}
For the era of radiation domination, $w=\frac{1}{3}$ followed by $a\propto \tau$, and for the era of matter domination, $w=0$ and $a\propto \tau^{2}$. Since the EoS before BBN ($\sim$ 1MeV) is still not clear, in the general scenario, we set $w$ as a free parameter, which represents the effective average EoS in this era.

The transfer function caused by general EoS is \cite{Zhao_2011}
\begin{equation}
 \label{transfer_function_of_EoS}
   T_{\text{EoS}}(k, \tau_{0})=\Big(\frac{a_{\text{b}}}{a(\tau_{0})}\Big)^{\beta}\Big(\frac{H_{\text{b}}}{k}\Big)^{\beta}\ \ \text{for}\ k> k_{\text{b}},
    \end{equation}
where $a_{\text{b}}$ is the scale factor at BBN epoch (here we chose the epoch at $T=1$MeV without losing generality \cite{Early_Universe_1990}) and $H_{\text{b}}$ is the Hubble parameter at BBN epoch. The lower boundary $k>k_{\text{b}}$ is determined by the fact that only those modes entered before $T_{\text{b}}$ are affected by unusual EoS.

Note that the sensitive band of PTA, i.e. $10^{-9}\sim10^{-7}$ Hz, corresponds to $52\sim 3.9\times10^3$ MeV, which is above the lower boundary of BBN energy scale, therefore unusual EoS could affect the detection of  PTAs. In this sensitive frequency range, the tensorial index $n_t$ and the EoS $w$ are degenerate. To see this clearly, we put the transfer function of cosmic expansion $T_z(k, \tau_0)$ and that of unusual EoS $T_{\rm EoS}(k, \tau_0)$ into the total transfer function. In the PTA sensitive band, $k\tau_0\gg0$, then from the spectrum (\ref{expanded_spectrum_of_energy_density}) we find $\Omega_{\rm gw}(k)\propto k^{n_t-\frac{2(1-3w)}{1+3w}}$.
The observations are influenced by the strength of RGWs, so they can merely determine the total tilt of $\Omega_{\rm gw}(k)$, which is a combination of $n_t$ and $w$.

\begin{figure}[!htb]
  \begin{center}
    \includegraphics[width=8.5cm]{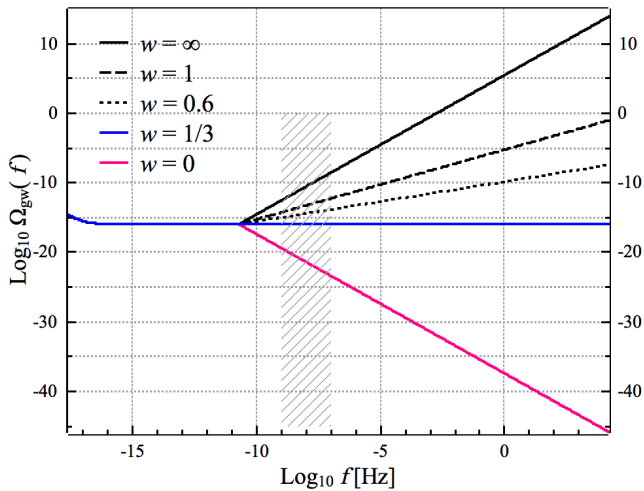}
   \end{center}
 \caption{\small The dependence of $\Omega_{\rm gw}(f, \tau_0)$ on the EoS before BBN. Five kinds of EoS, including $w=\infty$, 1, 0.6, 1/3 and $0$, are considered. Here we choose $r=0.1, n_{t}=0$ and adopt the cosmological parameters mentioned before. As the effect of unusual EoS changes the spectrum by several orders of magnitude, we do not include the relatively small effects caused by cosmic phase transitions $T_{\rm PT}(k, \tau_0)$ and free-streaming particles $T_\nu(k, \tau_0)$ here, but the damping effect of cosmic expansion $T_z(k, \tau_0)$ is still incorporated into the total transfer function.  Sensitive band of PTAs is the shaded area in the figure.}
 \label{spectrum_energy_density_EoS}
 \end{figure}

Fig.\ \ref{spectrum_energy_density_EoS} shows the spectrum of energy density undergoes the cosmic expansion and different EoS. For the unusual EoS with $w > 1/3$, the energy density spectrum of RGWs is strongly intensified, while the matter-like EoS ($w=0$) damps the RGWs greatly. It is clear that, when general EoS is considered, the spectrum at higher frequency is amplified or damped more than that at lower frequency.

Tab.\ \ref{Spectrum_at_particular_f_EoS} lists the exact strength of $\Omega_{\rm gw}$ at three different frequencies. For the extremely stiff EoS with $w=\infty$ ($w=1$) , the strength could be amplified at least $10^{4}$ ($10^{2}$) times for the sensitive band of PTAs. For the mild EoS $w=0.6$ \cite{Boyle_buonanno_2008,LIGO_correlation}, $\Omega_{\rm gw}$ could be amplified $9$ times at $10^{-9}$ Hz, and as far as $130$ times at $10^{-7}$ Hz. While a matter-like EoS ($w=0$) may damp the strength by four orders of magnitude.

\begin{table}[!htp]
 \begin{center}
   \begin{tabular}{c @{\extracolsep{1.3em}} c c c}
   \hline\hline
     $\gape w$ &  $\gape \Omega_{\rm gw}(10^{-9}$Hz) & $\gape \Omega_{\rm gw}(10^{-8}$Hz)  & $\gape \Omega_{\rm gw}(10^{-7}$Hz)\\
      \hline
   $\infty$ & $\gape 2.6\times10^{-13}$  & $\gape 2.5\times10^{-11}$  & $\gape 2.5\times10^{-9}$\\
    $1$  & $5.0\times10^{-15}$  & $5.0\times10^{-14}$  & $5.0\times10^{-13}$  \\
    $0.6$ & $9.2\times10^{-16}$  &  $3.4\times10^{-15}$   & $1.3\times10^{-14}$ \\
    $1/3$  & $9.8\times10^{-17}$  & $9.8\times10^{-17}$ & $9.8\times10^{-17}$  \\
    $0$   & $3.8\times10^{-20}$  & $3.8\times10^{-22}$ & $3.8\times10^{-24}$\\
    \hline
   \end{tabular}
   \end{center}
  \caption{\small The strength of $\Omega_{\rm gw}(f, \tau_0)$ at $f=10^{-9}, 10^{-8}$ and $10^{-7}$ Hz for EoS with $w=\infty, 1, 0.6, 1/3$ and 0 respectively. In the calculation, we use  $r=0.1$ and $n_{t}=0$. The cosmological parameters are the same as those mentioned at the beginning of this section.}
  \label{Spectrum_at_particular_f_EoS}
 \end{table}

\section{Constraints on RGWs by current pulsar timing arrays}

From Fig.\ \ref{spectrum_of_energy_density_all_effects} and Fig.\  \ref{spectrum_energy_density_EoS}, we have seen that many damping effects such as cosmological redshift, phase transition of QCD,  the free-streaming dark fluids and the unusual EoS are well in the sensitive band of PTAs. Thus PTAs provide the good probe into these interesting physics. Now let us turn to the detection of RGWs through PTAs.

RGWs could weakly stretch or suppress the space, leading to the change of time of arrival for electro-magnetic pulses from pulsars. Difference of time of arrival could be accumulated to form considerable timing residuals. By analyzing the timing residuals carefully it is possible to detect RGWs.

In practice, the characteristic strain $h_c(f)$ is more convenient for analysis and can be related to $\Omega_{\rm gw}$ by \cite{Maggiore_2000}
\begin{equation}
  \label{definition_of_hcf}
  \Omega_{\rm gw}(f,\tau_0)=\frac{2\pi^2}{3H^2(\tau_0)}f^2h_c^2(f).
   \end{equation}
So if the spectrum of energy density (\ref{expanded_spectrum_of_energy_density})
 is given, the characteristic strain can be determined theoretically.

When the spectrum is unknown, it is helpful to postulate a parameterized power-law form as follows \cite{NANO_2015, EPTA_2015},
\begin{equation}
  \label{parametrize_hcf}
   h_{c}(f)=A\Big(\frac{f}{f_{\text{yr}}}\Big)^{\alpha},
   \end{equation}
where $A$ is the amplitude of $h_{c}(f)$ at $f_{\rm yr}=1/\rm yr=3.171\times10^{-8}$ Hz, and $\alpha$ is the power index.

\begin{figure}[!htb]
  \begin{center}
    \includegraphics[width=8.5cm]{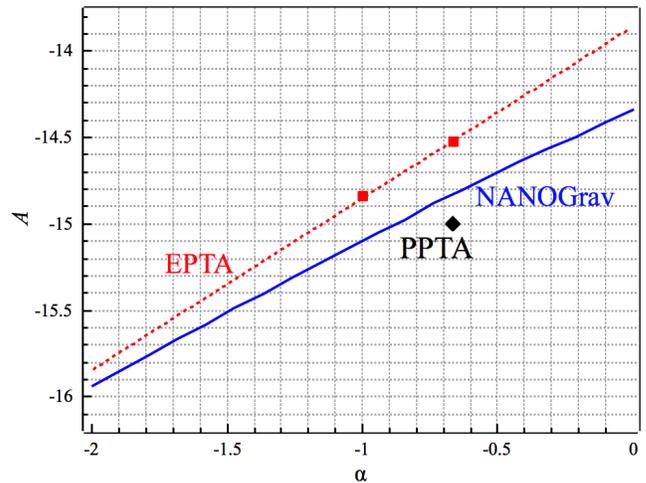}
    \end{center}
   \caption{\small The current 2$\sigma$-constraints on strain amplitude $A$ from three PTA projects: NANOGrav, EPTA and PPTA. The NANOGrav data are obtained through figure 3 in \cite{NANO_2015}. The two red squares
($\alpha=-1, A=1.4\times10^{-15}$) and $(\alpha=-2/3, A=3\times10^{-15})$ are constraints from EPTA by use of Bayesian method \cite{EPTA_2015}. The dotted line is obtained by linearly fitting the squares. The black diamond ($\alpha=-2/3, A=1.0\times10^{-15}$) comes from PPTA \cite{Parkes_2013}.}
   \label{constraints_on_a}
  \end{figure}

Fig.\ \ref{constraints_on_a} shows the current constraints on $A$ from PPTA  \cite{Parkes_2013}, EPTA \cite{EPTA_2015} and NANOGrav \cite{NANO_2015}. From Fig.\ \ref{constraints_on_a}, we see that the result of PPTA is the most stringent one (about a factor of 1.5 better than NANOGrav at $\alpha=-2/3$), but there is only one data point (the black diamond in the figure), so in the rest part of this paper we would concentrate on results from NANOGrav and EPTA to cover more possible indices. In a wide range of $\alpha$, NANOGrav gives a tighter constraint than EPTA for all $\alpha\in[-2, 0]$ and the advantage becomes obvious as $\alpha$ increases. For example, at $\alpha=-1$, NANOGrav gives $A<8.1\times10^{-16}$, which is only a factor of 1.7 better than $1.4\times10^{-15}$ of EPTA. When $\alpha$ increases to $-0.6$, NANOGrav gives $A<1.6\times10^{-15}$, a factor of 2.2 better than $3.5\times10^{-15}$ of EPTA.

\subsection{Constraints on RGWs by current PTAs considering cosmic expansion only}

When upper limits on $A$ for different indices are given, we can convert $h_{c}(f)$ into $\Omega_{\rm gw}(f)$ through (\ref{definition_of_hcf}) and compare it with the theoretical results in (\ref{expanded_spectrum_of_energy_density}) to give limits on the inflationary parameters $r$ and $n_t$.

We should firstly find out the relation between the two power indices $\alpha$ and $n_t$. Because different damping effects alter the spectrum differently (see Fig.\ \ref{spectrum_of_energy_density_all_effects}), the relation depends on the damping effects. Here we consider the spectrum damped by cosmic expansion only. In this case, the transfer function (\ref{separation_of_transfer_function}) has only one term $T_z(k, \tau_0)$ from cosmic expansion (\ref{transfer_function_redshift}). For waves with frequencies $2\pi f\gg k_{\rm eq}$, we find the simple relation between $\alpha$ and $n_t$
\begin{equation}
  \label{alpha_n_t_relation}
   n_t=2\alpha+2,
    \end{equation}
and the relation between $A$ and $r$

\begin{equation}
 \label{expresson_of_r_redshift}
  r=\frac{32\pi^2A^2k_{\rm eq}^2\tau_0^4}{45A_R(k_0)\Omega_{\rm m}^2}\Big(\frac{k_0}{2\pi}\Big)^{n_t}\Big(\frac{1}{f_{\rm yr}}\Big)^{n_t-2}.
  \end{equation}
Inserting the parameters in Section \ref{section2} and the data in Fig.\ \ref{constraints_on_a}, we obtain the upper limits on $r$ for different $n_t$.

\begin{figure}[!htb]
  \begin{center}
    \includegraphics[width=8.5cm]{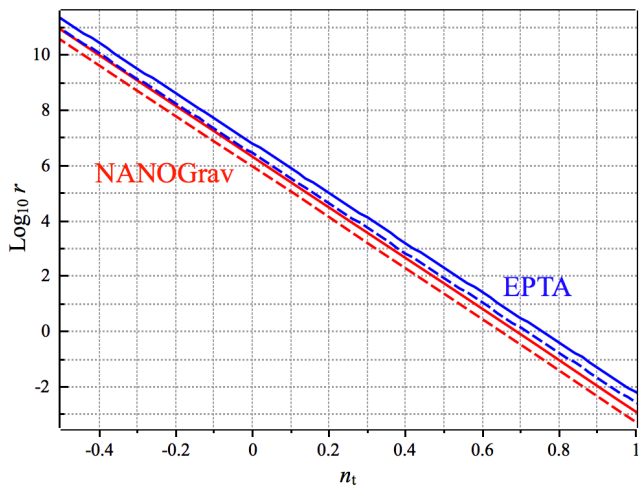}
   \end{center}
   \caption{\small $2\sigma$-constraints on the $r-n_t$ space by NANOGrav \cite{NANO_2015} and EPTA\cite{EPTA_2015}. The solid lines are the constraints for the case including transfer function of cosmic expansion $T_z(k, \tau_0)$ and that of cosmic phase transitions $T_{\rm PT}(k, \tau_0)$, while the dotted lines contain $T_z(k, \tau_0)$ only.}
  \label{constraints_on_r_nano_epta}
 \end{figure}

Fig.\ \ref{constraints_on_r_nano_epta} shows the constraints on $r-n_t$ space. The  red dashed and the blue dashed line are the cases considering $T_z(k, \tau_0)$ only. At $n_t=0$, the limit from NANOGrav is $8.5\times10^5$, which is a factor of $3.1$ more stringent than  $2.7\times10^6$ from EPTA. Note that \cite{EPTA_2015} gives a limit $2.5\times10^6$ for the same EPTA data. The small deviation occurs because we adopt new cosmological parameters, which are slightly different from those used in \cite{EPTA_2015}. As $n_t$ increases, the constraints become more stringent. At $n_t=0.8$, the constraint of $r$ from NANOGrav is $0.04$, while upper limit of EPTA is $0.17$ (see Table\ \ref{NANOGrav_EPTA_a_r_nt}). For $n_t<0$ both constraints become very loose. In Table\ \ref{NANOGrav_EPTA_a_r_nt}, we also list the results for $n_t=-0.4$.

For $r=0.1$, when only redshift is considered, Zhao et al \cite{Zhao_et_al_2013} found $n_t=0.90$ for  NANOGrav \cite{NANO_2013} and $n_t=0.88$ for EPTA \cite{EPTA_2011}, based on the previous observations. While we found that, for $r=0.1$, latest NANOGrav data \cite{NANO_2015} follows $n_t=0.76$ and EPTA data \cite{EPTA_2015} follows $n_t=0.83$. Both constraints become tighter due to the update of NANOGrav and EPTA data.

\subsection{Constraints on RGWs by current PTAs considering phase transitions and neutrinos}

Although the damping effects beyond cosmic expansion $T_z(f, \tau_0)$, such as cosmic phase transitions $T_{\rm PT}(k, \tau_0)$ and the relativistic free-streaming gases $T_{\nu}(k, \tau_0)$, have been well studied before, few authors include these  damping effects when they constrain the inflationary parameters (see \cite{Zhao_et_al_2013} for example). As we mentioned before, cosmic phase transitions could damp the spectrum $\Omega_{\rm gw}(f, \tau_0)$ as much as 60\% in the sensitive band of PTAs (see Fig.\ \ref{spectrum_of_energy_density_all_effects}). Thus, to make accurate detection, it is inevitable to consider these damping effects. In this subsection, we will discuss the constraints on the inflationary parameters $r$ and $n_t$ when additional damping effects $T_{\rm PT}(k, \tau_0)$ and $T_{\nu}(k, \tau_0)$ are included in the transfer function.

There are two points that deserve our attention. Firstly, it is very clear that neutrinos could not affect any detection approached by PTAs, as neutrinos only damp the spectrum in the range of $10^{-16}\sim 10^{-10}$ Hz, which is well beyond the detecting ability of PTAs. Secondly, $T_{\rm PT}(k, \tau_0)$ makes the spectrum deviate from power law, so, strictly speaking, the postulation (\ref{parametrize_hcf}) does not hold. But $T_{\rm PT}(k, \tau_0)$ is on the order of 0.1, which could only change the spectrum by a small amount, so in the meaning of perturbation, we could still apply the power law and attribute the small deviation from power law to the weak dependence of $A$ on the frequency.  Then the relation (\ref{alpha_n_t_relation}) between $\alpha$ and $n_t$ remains, while the expression of $r$ should be modified.

When we include cosmic phase transitions, the transfer function $T(k, \tau_0)$ will have another term $T_{\rm PT}(k, \tau_0)$ (\ref{transfer_function_of_phase_transition}) besides  $T_z(k, \tau_0)$ (\ref{transfer_function_redshift}), so the expression of $r$ becomes
\begin{equation}
 \label{scalar_to_tensor_ration_redshift_pt}
   r=\frac{32\pi^2A^2k_{\rm eq}^2\tau_0^4}{45A_R(k_0)\Omega_{\rm m}^2}\Big(\frac{k_0}{2\pi}\Big)^{n_t}\Big(\frac{1}{f_{\rm yr}}\Big)^{n_t-2}\times\frac{1}{T_{\rm PT}^2(k, \tau_0)}.
  \end{equation}
As $T_{\rm PT}(k, \tau_0)\le1$, we can expect the upper limit on $r$ for fixed $n_t$ would increase after we include cosmic phase transitions. Specifically, numerical calculations give $T^2_{\rm PT}(k_{\rm yr}, \tau_0)=0.428$, then we could obtain a looser constraint on $r-n_t$ space and the upper limit on $r$ increases a factor of 2.3 for all indices.

Fig.\ \ref{constraints_on_r_nano_epta} shows the upper limit on $r-n_t$. The red solid and the blue solid line are upper limits where $T_{\rm PT}(k, \tau_0)$ is included. At $n_t=0$, NANOGrav \cite{NANO_2015} gives upper limit  $2.0\times10^6$, and EPTA \cite{EPTA_2015} gives $6.4\times10^6$. Both are indeed raised by a factor of 2.3 compared to the limit without  $T_{\rm PT}(k, \tau_0)$. At $n_t=0.8$, the constraint from NANOGrav is $r<0.09$ and that given by EPTA is $r<0.39$. We also list the results at $n_t=-0.4$ in Table\ \ref{NANOGrav_EPTA_a_r_nt}.

\begin{table}[!htb]
\centering
\begin{tabular}{c @{\extracolsep{1.3em}} c c c c}
\hline\hline
 \multirow{2}*{$n_t$} & \multicolumn{2}{c}{Include $T_z$ only} & \multicolumn{2}{c}{Include $T_z$ and $T_{\rm PT}$} \\
  \Xcline{2-3}{0.4pt}\Xcline{4-5}{0.4pt}
  & $r_{\rm nano}$ & $r_{\rm epta}$  & $r_{\rm nano}$ & $r_{\rm epta}$ \\
 \hline
  $\gape -0.4$ & $\gape 4.1\times10^9 $ &  $\gape 1.1\times10^{10}$ & $\gape 9.6\times10^{9}$  & $\gape 2.6\times10^{10}$ \\
  $0$ & $8.5\times10^5$ & $2.7\times10^6$  & $2.0\times10^6$  & $6.4\times10^6$ \\
 $0.8$ & $0.04$ & $0.17$  & $0.09$ & $0.39$\\
 \hline
 \end{tabular}
 \caption{Constraints on upper limit of $r$ for different $n_t$ by use of NANOGrav \cite{NANO_2015} and EPTA \cite{EPTA_2015}. The second and third column are the limits when cosmic expansion $T_z(k, \tau_0)$ is the only transfer function, while in the last two columns cosmic phase transition $T_{\rm PT}(k, \tau_0)$ is added.}
\label{NANOGrav_EPTA_a_r_nt}
\end{table}

\subsection{Constraints on RGWs from other observations}

 \subsubsection{Constraints from BBN}
  \label{Constraints from BBN}

Besides PTA projects, BBN can also give upper limit on $r-n_t$ space. BBN precisely predicts the abundance of light elements in the Universe. Nucleosynthesis happens efficiently when the collision rate of protons and neutrons are much higher than the Hubble parameter at that moment. To obtain enough abundance of H and He, the expansion rate of the Universe should not be too large, otherwise the plasma soup would be diluted too much to produce effective reaction. Expansion rate of the Universe or Hubble parameter $H$ depends on the energy density of radiation at that moment, thus the abundance of light elements could provide constraints on the total energy of radiation including RGWs.

By use of the effective number of  neutrino species $N_{\rm eff}$, BBN could set an upper limit on the energy of RGWs \cite{Maggiore_2000}
\begin{equation}
 \label{BBN_constraining}
   \int_{k_{\rm low}}^{k_{\rm up}} \Omega_{\rm gw}(k, \tau_0)h^2~{\rm d}\ln k\leq 5.6\times10^{-6}(N_{\rm eff}-N_\nu),
   \end{equation}
where $N_\nu$ is the number of neutrino species, while $N_{\rm eff}$ is the effective number of neutrino  species. For standard model in particle physics, considering the non-instantaneous decoupling of neutrinos, $N_\nu=3.046$ \cite{Mangano_2002}. Recent combined observations from BBN and primordial mass fraction of $^4$He and the abundance of deuterium give $N_{\rm eff}=3.41$ with $95\%$ confidence \cite{Cyburt_2015}, then $\int \Omega_{\rm gw} h^2~{\rm d}\ln k\le 2.04\times 10^{-6}$. The lower boundary of the integration is determined by counting the RGWs that entered horizon at the BBN epoch only, while the upper boundary is determined by quantum limit. For the very short wavelength (i.e. high frequency) portion, the ultraviolet divergences is avoided by considering the Parker's adiabatic theorem \cite{parker}, which states that, during a transition between expansion epochs with a characteristic time during $\Delta t$, the gravitons created will be suppressed for wavenumbers $k>1/\Delta t$. Here we follow \cite{Miao_Zhang_2007} to choose $f_{\rm up}=k_{\rm up}/2\pi=10^{10}$ Hz by assuming the energy scale for the inflation is around $10^{16}$ GeV and $k_{\rm low}=k(T=1{\rm MeV})$ which can be obtained from (\ref{transformation_from_t_to_k}).

\begin{figure}[!htb]
 \begin{center}
  \includegraphics[width=8.5cm]{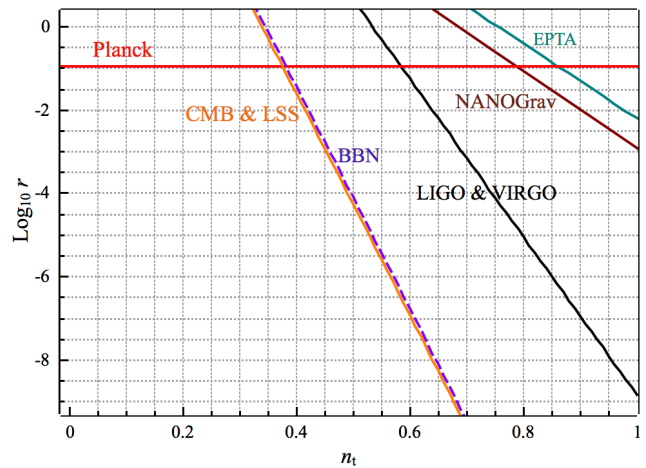}
  \end{center}
 \caption{\small Constraints on the inflationary parameter space $r-n_t$ by use of data from observations of CMB (Planck \cite{Planck_parameters_2015} and the joint analysis of CMB and LSS (matter power spectrum)  \cite{Pagano_2015}), BBN  \cite{Cyburt_2015}, LIGO \& Virgo \cite{LIGO_correlation}, EPTA and NANOGrav \cite{EPTA_2015, NANO_2015}. Note that the line stands for BBN is only a bit higher than that for CMB \& LSS, so we use purple dashed line to represent the result of BBN and orange solid line for CMB \& LSS. Here all the constraints have considered $T_z(k, \tau_0)$, $T_{\rm PT}(k, \tau_0)$ and $T_\nu(k, \tau_0)$.}
 \label{BBN_LIGO_CMB_phase_transition}
 \end{figure}

Inserting the transfer function $T_z(k, \tau_0)$ (\ref{transfer_function_redshift}) caused by cosmic expansion and $T_{\rm PT}(k, \tau_0)$ by cosmic phase transitions (\ref{transfer_function_of_phase_transition}) into the spectrum of energy density $\Omega_{\rm gw}(k, \tau_0)$ in (\ref{expanded_spectrum_of_energy_density}), we can integrate the spectrum numerically to obtain the upper limit on the inflationary parameter space $r-n_t$. The result is shown by the purple dashed line in Fig.\ \ref{BBN_LIGO_CMB_phase_transition}.

 \subsubsection{Constraints from CMB and matter power spectrum}
 \label{constraints from cmb}

Due to the precise observations of temperature and polarization anisotropies, in the modern cosmology, CMB becomes one of the most important tools to constrain various cosmological phenomena/processes including RGWs. RGWs affect the CMB fluctuations by two different ways.

Firstly, RGWs, as well as density perturbations, are the sources which generate the CMB anisotropy power spectra, including the temperature anisotropy auto-correlation spectrum (TT), E-mode polarization auto-correlation spectrum (EE), B-mode polarization auto-correlation spectrum (BB), and the cross-correlation of temperature anisotropy and E-mode polarization (TE). Although the definite signal of RGWs has not yet been found in the CMB power spectra, based on the recent observations of TT, TE and EE data, Planck team gives the quite tight constraint on the tensor-to-scalar ratio $r<0.11$ \cite{Planck_parameters_2015}, which is nearly independent of the spectral index $n_t$ \cite{huang}. A similar constraint $r<0.12$ is also obtained from the joint analysis of BICEP2/{Keck Array} and {Planck} B-mode polarization data \cite{Bicep2_planck_2015}.

 Secondly, RGWs affect the growth of density perturbation and the CMB anisotropy power spectra by contributing extra energy density to the total energy density or, equivalently, by changing the expansion rate of the Universe. The weak interaction of RGWs with matter makes them similar to massless neutrinos. So constraints on additional neutrinos from CMB and large-scale structure (LSS) observations could be applied to RGWs \cite{Smith_2006}. With $95\%$ confidence level, CMB combined with matter power spectra gives $N_{\rm eff}-N_\nu<0.33$ \cite{Pagano_2015}. Again, we can insert this number into (\ref{BBN_constraining}) to yield another constraints on RGWs. We obtain $\int \Omega_{\rm gw} h^2 ~{\rm d}\ln k\le 1.85\times 10^{-6}$, but the lower boundary of integration is much lower than that of BBN. We adopt $f_{\rm low}=10^{-15}$ Hz \cite{Smith_2006} to do the integration. The corresponding constraint on $r-n_t$ is shown in Fig.\ \ref{BBN_LIGO_CMB_phase_transition} (the orange solid line).

\subsubsection{Constraints from gravitational-wave detectors}

The ground-based interferometer could directly detect RGWs in the range $10^1\sim 10^3$ Hz by analyzing the interference stripes caused by gravitational waves in the laser beams. The latest constraints on the stochastic background is from joint observations of LIGO and Virgo. In frequency range $41.5-169.25$ Hz, LIGO and Virgo correlations give an upper limit on energy density spectrum $\Omega_{\rm gw}(k, \tau_0)<5.6\times10^{-6}$ for $n_t=0$, while in the range $600-1000$ Hz $\Omega_{\rm gw}(k, \tau_0)<0.14$ for $n_t=3$ (there are the other two sets of constraints at $n_t=0$ and $3$, but they are much looser, so we ignore them). Both limits are of 95\% confidence \cite{LIGO_correlation}.

 We can use the LIGO \& Virgo data to estimate the upper limits of inflationary parameters $r$ and $n_t$. Firstly, we convert the limits on $\Omega_{\rm gw}(k, \tau_0)$ at $\alpha=0$ and $3$ into strain amplitude $A$ through (\ref{definition_of_hcf}), then use (\ref{scalar_to_tensor_ration_redshift_pt}) to obtain limits on $r$ and $n_t$ at these two indices. Finally, we linearly fit the two points to cover the range in $n_t\in(0,1)$, which is shown in Fig.\ \ref{BBN_LIGO_CMB_phase_transition} (the black solid line).

\subsubsection{Comparison with different constraints}

Fig.\ \ref{BBN_LIGO_CMB_phase_transition} shows the constraints on $r-n_t$ space from different observations. We can see that at low index ($n_t<0.37$),  upper limit of $r$ from Planck is the most stringent and  constant constraint ($r<0.11$), while at high index ($n_t>0.37$), constraint by use of joint analysis of CMB and LSS becomes the most stringent and the upper limit decreases as $n_t$ increases, for example, at $n_t=0.5$, the upper limit is $10^{-4}$, but at $n_t=0.8$, the limit is $r<10^{-12}$.

We see that the constraints from BBN (the purple dotted line) is slightly looser than that from the joint data of CMB and LSS (the orange solid line). There are two reasons for it. Firstly, constraints on $N_{\rm eff}$ from BBN is a bit looser than that from the joint analysis (0.36 versus 0.33, see subsection \ref{Constraints from BBN} and \ref{constraints from cmb}.) Secondly, the integration range of BBN ($f_{\rm low}\sim 10^{-10}$ Hz) is larger than that of the combination of CMB and LSS ($f_{\rm low}\sim10^{-15}$ Hz), thus less energy could be added to the integration, making the tensor index larger.

Although the direct constraints from EPTA and NANOGrav have significantly improved recently, they are still much weaker than those from CMB, matter power spectra and BBN. So are LIGO and Virgo. Therefore, the great advancement is still needed to put tighter constraints. Advanced LIGO and future PTAs like FAST and SKA are indeed quite necessary.

\section{Detection of RGWs by the future observations}

In this section we will forecast the detections of RGWs by the potential PTA observations, including the FAST and SKA projects. In addition, as we know, the damping effects caused by cosmological redshift, phase transition of QCD, the free-streaming dark fluid and the unusual EoS are all well in the sensitive band of PTAs. So, the future PTAs also provide a good chance to study these interesting physics, which will also be investigated in this section.

In order to estimate detection abilities of the future PTAs, we will use time of arrival to establish the signal-to-noise ratio (SNR) of observations. When the spectrum of energy density $\Omega_{\rm gw}$ (\ref{expanded_spectrum_of_energy_density}) is given, an important problem follows: What confidence level can we achieve when we come up with new PTA projects to detect RGWs? The key lies on the timing residuals caused by RGWs and their correlations. RGWs could stretch or suppress the space weakly, leading to the change of time of arrival for electro-magnetic pulses from pulsars. Difference of time of arrival could be accumulated to form considerable timing residuals. The timing residuals are usually decomposed into two parts
\begin{equation}
 \label{timing_residuals}
   R^{i}(t_{k})= s^{i}(t_{k})+n^{i}(t_{k}),
    \end{equation}
where $R^{i}(t_{k})$ is the timing residual for $i$-th pulsar at time $t_{k}$, while $s^{i}(t_{k})$ and $n^{i}(t_{k})$ are the timing residuals contributed by RGWs and by noise. To make a detection, it is necessary to discriminate the contribution of RGW signal from that of noise. Let us consider the characteristic of $s^{i}(t_{k})$ and $n^{i}(t_{k})$ separately.

The genuine RGWs would induce common residuals in the data, so the correlation of $s^{i}(t_{k})$ gives \cite{Lee_bassa_2012}
\begin{equation}
    \langle s^{i}(t_{k}) s^{j}(t_{k'})\rangle =\sigma_{\text{g}}^{2}H(\theta_{ij}) \delta_{t_{k},t_{k'}},
    \end{equation}
where $\delta_{t_{k},t_{k'}}$ is Kronecker delta function and $H(\theta_{ij})$ is the Hellings-Downs curve, while $\sigma_{\text{g}}$ is root mean square (RMS) of timing residuals resulted from RGWs. The Hellings-Downs curve \cite{Hellings_downs_1983, Lee_et_al_2008} is
\begin{equation}
 H(\theta_{ij})=\frac{3}{2}x\ln x-\frac{x}{4}+\frac{1}{2}\Big(1+\delta(x)\Big), x\equiv\frac{1-\cos\theta_{ij}}{2}
  \end{equation}
where $\theta_{ij}$ is the angular separation between $i$-th and $j$-th pulsar. The RMS of timing residuals from RGWs is given by
\begin{equation}
  \sigma_{\text{g}}^{2}=\int^{f_{h}}_{f_{l}} \frac{\big|h_{c}(f)\big|^{2}}{12\pi^{2}f^{3}} df,
  \end{equation}
where $f_{l}=1/T$ and $f_{h}=1/(2\Delta t)$ are the lower and the higher boundary of sensitive band. $T$ and $\Delta t$ are the span of whole observation and the interval between two observations respectively. Now, if the spectrum of energy density (\ref{expanded_spectrum_of_energy_density}) is given, the RMS of RGW is determined.

As for the part of noise, we assume that they are white and the same for every pulsar in the PTA, then the correlation of contributions from the noise is simple \cite{Lee_bassa_2012}
\begin{equation}
  \langle n^{i}(t_{k}) n^{j}(t_{k'})\rangle =\sigma^{2}_{\text{n}} \delta_{ij}\delta_{t_{k}, t_{k'}},
    \end{equation}
where $\sigma_{\text{n}}$ is the RMS of timing residuals from white noise.

By use of the correlation of timing residuals and the whitening method, we could obtain the SNR for PTA \cite{Jenet_et_al_2005, Zhao_et_al_2013, Kuroyanagi_et_al_2013}
\begin{equation}
  \label{averageSNR}
  \langle {\rm SNR} \rangle =\sqrt{N}\Bigg\{ 1+\frac{\sum_{\Delta'}\Big[1+\Big(\frac{P_{\text{g}}(\Delta')}{P_{d}(\Delta')}\Big)^{2}\overline{H^{2}}\Big]}{\Big(\sum_{\Delta'}\frac{P_{\text{g}}(\Delta')}{P_{\text{d}}(\Delta')}\Big)^{2}\sum_{H}^{2}}\Bigg\}^{-\frac{1}{2}},
   \end{equation}
where brackets $\langle\ \rangle$ are ensemble average (hereafter when we say SNR we mean the expectation of SNR, or $\langle {\rm SNR}\rangle$). $N=n(n-1)/2$ is the number of pairs with $n$ being the number of pulsars. $\overline{H^{2}}$ and $\Sigma_{H}$ are the average of $H^{2}(\theta_{ij})$ and the variance of $H(\theta_{ij})$ respectively. For pulsars evenly distributed in the full sky, we have $\overline{H^{2}}=\Sigma^{2}_{H}=1/48$.

As for $P_{\text{g}}(\Delta')$, it is defined as the RMS in the $\Delta'$-th bin \cite{Jenet_et_al_2005}:
\begin{equation}
   \label{pgdelta}
   P_{\text{g}}(\Delta')=\int^{\frac{\Delta'+0.5}{T}}_{\frac{\Delta' -0.5}{T}} \frac{h_{c}^{2}(f)}{12\pi^{2}f^{3}}df,  \quad (\Delta'> 1)
   \end{equation}
where $\Delta'$ is the bin $\Delta$ whose signal is higher than noise, i.e. $P_{\text{g}}(\Delta)>\sigma_{\text{n}}^{2}/m$, with $m$ being the total number of  observations in the PTA project. For $\Delta'=1$ the lower boundary of integration in $P_{\text{g}}(\Delta')$ is $0.97/T$ \cite{Jenet_et_al_2005}. The total spectral density in a bin is the sum of contributions from signals and noises: $P_{\text{d}}(\Delta')=P_{\text{g}}(\Delta')+\sigma_{\text{n}}^{2}/m$.

We could see that, to calculate the SNR for a given spectrum of energy density, an observational plan should at least include the following information: the observation  span $T$, the number of pulsars $n$, the number of observations $m$ and the noise level.

For timing noise, due to the intrinsic instability in the pulsars and uncertainties in the telescopes, the noise is not guaranteed to be white and may have different value for different pulsars \cite{Lee_bassa_2012}.  In our analysis, for simplicity, we assumed white and equal noise level $\sigma_{\text{n}}$ for all pulsars in a given array.

\subsection{Potential PTA projects in the future}
\label{future_pta_projects}

 \label{Detection of RGWs in the accelerating Universe}
In this paper, we follow Zhao et al \cite{Zhao_et_al_2013} to consider the following future observations: the complete PPTA, FAST, SKA and optimal PTA.

The complete PPTA plans to observe 20 pulsars which have the same noise level of 100 ns over 5 years ($T=5$ yr, $\sigma_{\text{n}}=100$ ns, $n=20$, $m=250$), while current PPTA mainly depends on four pulsars \cite{Parkes_2013}. So the future PPTA could greatly improve the observation.

Another observation plan that deserves our attention is that from FAST, which would be the largest single dish in the world when it obtains the first light in 2016. Working with multi beams in the 70 MHz$-$3GHz frequency band, FAST is expected to discover $~$4000 Galactic pulsars and one tenth may be millisecond pulsars \cite{FAST_observational_plan_paper}. For the search for gravitational waves, FAST plans to run a series of similar observations ($T=5$ yr, $n=20$, $m=250$, we use FAST(20) to label this plan) but with much lower noise of only $30$ ns \cite{FAST_observational_plan_paper}, which may directly improve the value of SNR. Here we assume a second plan, FAST(40), to adequately cover the possibility of more pulsars in a longer duration of timing: $T=10$ yr, $n=40$, $m=500$.

In the 2020s, the establishment of the thousands of single dishes will erect the biggest radio array SKA and pulsar searching and timing will be part of the SKA scientific goals. Accounting for the large signal-collecting area, SKA could well survey the pulsars in the Milky Way in an unprecedented efficient way \cite{Kramer_M} and greatly facilitate the detection of gravitational waves. To estimate the capacity of SKA in detecting RGWs, we adopt a low-noise plan running for ten years ($T=10$ yr, $\sigma_{\text{n}}=50$ ns, $n=100, m=500$, we use SKA(100) to denote this plan) \cite{SKA_100_pulsars}. However, from a conservative view point, we would also consider a less ambitious one, SKA(40), with $T=10$ yr, $\sigma_{\text{n}}=50$ ns, $n=40, m=500$.

Prolonging time span allows PTAs to approach more low frequency signals, which can be much stronger than that of high frequency. So, accounting of these improvements, future PTAs may greatly increase their sensitivities. Zhao et al \cite{Zhao_et_al_2013} also considered the optimal PTA ($T=20$ yr, $\sigma_{\text{n}}=30$ ns, $n=200, m=1000$), which can reach an unprecedented sensitivity. For easy reference, we list these projects in Table\ \ref{potential_PTA}.

\begin{table}[!htb]
\centering
\begin{tabular}{c @{\extracolsep{2em}} c c c c}
\hline\hline
Potential PTA &  $T$/yr  & $\sigma_{\rm n}$/ns  &  $n$  &  $m$\\
\hline
Complete PPTA...  & 5  & 100  &  20  &  250 \\
FAST(20)............. & 5 &  30 &  20 & 250 \\
FAST(40)............. & 10 &  30 &  40 & 500 \\
SKA(40)...............  & 10 & 50 & 40 &500 \\
SKA(100).............  & 10 & 50 & 100 &500 \\
Optimal PTA....... & 20  & 30 & 200 & 1000\\
\hline
\end{tabular}
\caption{The parameters of potential PTA projects: Time span $T$, noise level $\sigma_{\rm n}$, number of observed pulsars $n$ and the total number of observations $m$.}
\label{potential_PTA}
\end{table}

For the known damping factor, if $r$ and $n_t$ are given, the spectrum (\ref{expanded_spectrum_of_energy_density}) is determined. Then, we use (\ref{definition_of_hcf}) to find the characteristic strain $h_c(f)$ and obtain $P_{\rm g}(\Delta')$ through (\ref{pgdelta}) for known PTA observations. Finally, we can calculate the SNR for these observations through (\ref{averageSNR}).

\begin{figure}[!htb]
  \begin{center}
   \includegraphics[width=8.5cm]{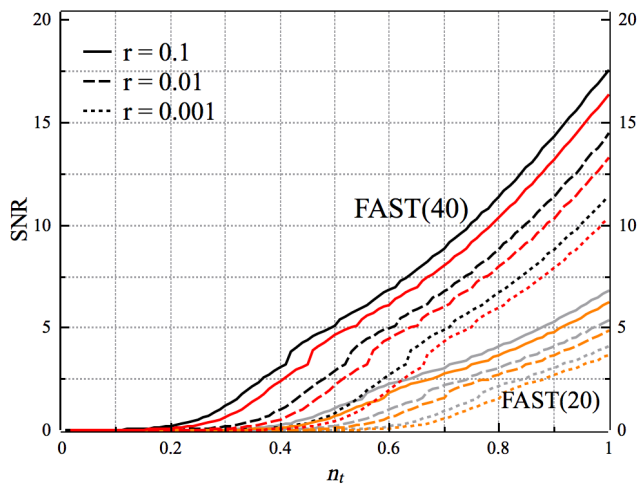}
    \end{center}
   \caption{\small The SNR of FAST(20) and FAST(40) when different damping effects are considered. For FAST(40), the black lines consider the effect of cosmic expansion $T_z(k, \tau_0)$ only, while the red ones include cosmic phase transitions $T_{\rm PT}(k, \tau_0)$. For FAST(20), the grey lines consider $T_z(k, \tau_0)$ only and the orange ones add $T_{\rm PT}(k, \tau_0)$. In each project, from up to down, the solid, dashed and dotted lines are for the case of $r=0.1, 0.01, 0.001$ respectively.}
   \label{snr1_under_phase_transition}
  \end{figure}

Note that without RGW signals, the timing residuals  (\ref{timing_residuals}) will be Gaussian distributed, thus ${\rm SNR}=2$ represents 95\% confidence level. In the calculation, we would like to obtain the constraints on $r-n_t$ space with 95\% confidence for the potential observations. This is possible, because for a given observational plans, when ${\rm SNR}$ is fixed, $r$ and $n_t$ are constrained by the relation (\ref{averageSNR}). So we firstly insert the spectrum with free parameters $r$ and $n_t$ into (\ref{definition_of_hcf}) to obtain $P_{\rm g}(\Delta')$ through (\ref{pgdelta}), then set ${\rm  SNR}=2$ to give upper limits on $r$ for different $n_t$ with 95\% confidence level.

\subsection{Detecting  RGWs in the accelerating Universe}

As the first step, we only consider transfer function due to cosmic expansion $T_z(k, \tau_0)$ in (\ref{transfer_function_redshift}) and assume a normal radiation EoS $w=1/3$ before BBN in this section.

Fig.\ \ref{snr1_under_phase_transition} show the SNR of FAST in the range $n_t\in[0, 1]$. The black and grey lines are the SNR including damping effect from $T_z(k, \tau_0)$ only. Our results for FAST(20) are the same as those in \cite{Zhao_et_al_2013}. At $n_t=0$, SNR vanishes for both FAST plans, thus no RGWs can be detected. However, as $n_t$ increases, the SNR increases a lot. If we fix $n_t=0.8$, when $r=0.1$, RGWs produce a significant detection with SNR$=4$ for FAST(20) and 11.5 for FAST(40).

It is obvious that smaller $r$ leads to lower SNR. For example, at $n_t=0.8$, when $r$ decreases from 0.1 to 0.001, SNR jumps from 11.5 to 6.5 for FAST(40), and from 4 to 2 for FAST(20), thus greatly weaken the detection.

\begin{figure}[!htb]
  \begin{center}
   \includegraphics[width=8.5cm]{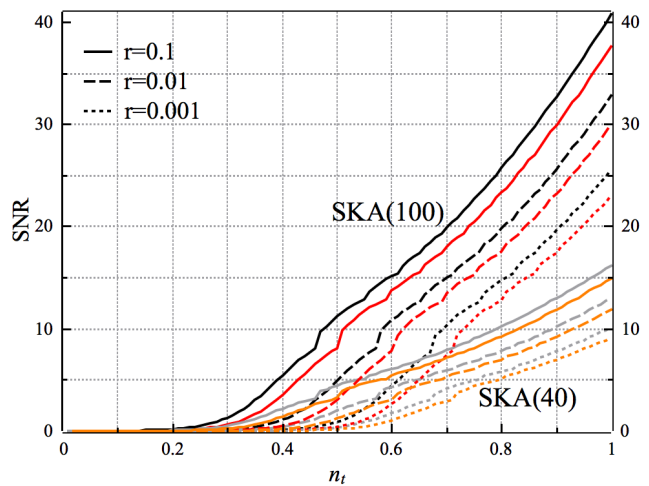}
    \end{center}
   \caption{\small The SNR of SKA(40) and SKA(100) when $T_z(k, \tau_0)$ and $T_{\rm PT}(k, \tau_0)$ are considered. The color is similar to Fig.\ \ref{snr1_under_phase_transition}.}
   \label{snr2_under_phase_transition}
  \end{figure}

The trends described above also hold for SKA, as we could see from Fig.\ \ref{snr2_under_phase_transition}, which shows the SNR for two SKA plans with 40 and 100 pulsars. Since the only difference between SKA(40) and SKA(100) is the number of pulsars (see Table.\ \ref{potential_PTA}), we conclude that increasing the scale of arrays could greatly raise the detection ability. Lower noise level also increases the possibility to make a detection, therefore, FAST(40) will performe better than SKA(40). For example, at $n_t=0.8$, for $r=0.1$, FAST(40) gives SNR=11.5, which is a bit higher than 10 given by SKA(40).

\begin{figure}[!htb]
 \begin{center}
  \includegraphics[width=8.5cm]{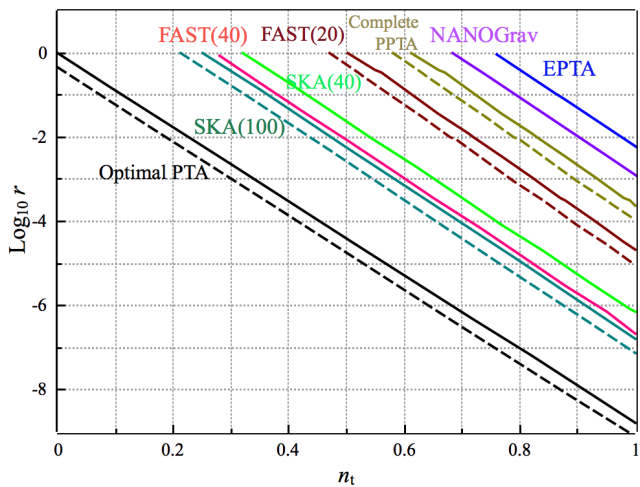}
   \end{center}
\caption{\small The constraints on $r-n_t$ space with 95\% confidence by future PTAs described in Table.\ \ref{potential_PTA}: Complete PPTA, FAST, SKA and Optimal PTA. Dashed lines are the constraints considering cosmic expansion $T_z(k, \tau_0)$ only, while solid lines consider both $T_z(k, \tau_0)$ and cosmic phase transitions $T_{\rm PT}(k, \tau_0)$. The lines of NANOGrav and EPTA are copied from Fig.\ \ref{BBN_LIGO_CMB_phase_transition} for comparison. Here, for clarity, we only plot the solid lines for FAST(40) and SKA(40).}
   \label{impact_of_pt_on_ptas}
  \end{figure}

In Fig.\ \ref{impact_of_pt_on_ptas}, we plot the constraints on $r-n_t$ space by potential observations. The dashed lines are the cases only considering $T_z(k, \tau_0)$. The current limits from NANOGrav and EPTA are also plotted here to make comparison. We could see, when only $T_z(k, \tau_0)$ is considered, in the whole range $n_t\in[0,1]$, for the upper limits on $r$, complete PPTA could be about $10$ times more stringent than current NANOGrav and about $50$ times better than current EPTA, while FAST(20) is $11$ times more stringent than complete PPTA and SKA(100) is 137 times better than FAST(20). The most stringent limit would come from the optimal PTA, which is about $107$ times better than SKA(100).

Although both SKA plans could defeat the FAST(20), the advantages of SKA would be greatly challenged by FAST(40), which is a factor of 2.5 better than SKA(40) and only a factor of 1.6 weaker than SKA(100). The good competence of FAST(40) comes from the low timing noise, as indicated by Table.\ \ref{potential_PTA}. As SKA(100) and FAST(20) are the upper and lower limits of the four plans of SKA and FAST, we would concentrate on these two observations in the rest part of this paper. However, due to the similar resolution of FAST(40) and SKA(100) in the $r-n_t$ space, the results for SKA(100) also hold for FAST(40) with good accuracy.

At $n_t=0$, upper limit on $r$ given by optimal PTA is $0.44$, which is the best constraints. At $n_t=0.8$, upper limit given by complete PPTA is $0.008$, while FAST(20) gives $0.0007$ and SKA(100) and the optimal PTA are much more stringent.

For $r=0.1$, the optimal PTA gives $n_t<0.08$, while for SKA(100) the upper limit is $0.32$, for FAST(20) it is $0.58$, and for complete PPTA, it is $0.69$. If $r=0.01$, the optimal gives limit $n_t<0.18$, SKA(100) gives $n_t<0.44$, and FAST(20) gives $n_t<0.68$. So both SKA(100) and optimal PTA could give fairy good constraints on the inflation models.

\subsection{Effects of phase transitions and neutrinos}

In the previous work \cite{Zhao_et_al_2013}, we only calculated the SNR for the RGWs damped by cosmic expansion $T_z(k, \tau_0)$, which is briefly reviewed in the previous subsection. However as we mentioned before, cosmic phase transitions, especially the phase transition of QCD, damp the spectrum by amount of 60\% at $f>10^{-9}$ Hz, which is well in the sensitive band of PTAs. Relativistic free-streaming neutrinos could also damp the spectrum by about 40\% in the range of $10^{-16}\sim10^{-10}$ Hz. Although the behavior of neutrinos has no effect on PTAs, they play an important role in the constraints by use of combined CMB and matter power spectra. Therefore effects from cosmic phase transitions and neutrinos deserve attention.

In this subsection, we shall take into account the damping effects caused by cosmic phase transitions $T_{\rm PT}(k, \tau_0)$ in (\ref{transfer_function_of_phase_transition}) and neutrinos $T_\nu(k, \tau_0)$ in (\ref{damping_factor_of_neutrinos}). Here we also assume a normal EoS with  $w=1/3$.

\begin{figure}[!htb]
 \begin{center}
  \includegraphics[width=8.5cm]{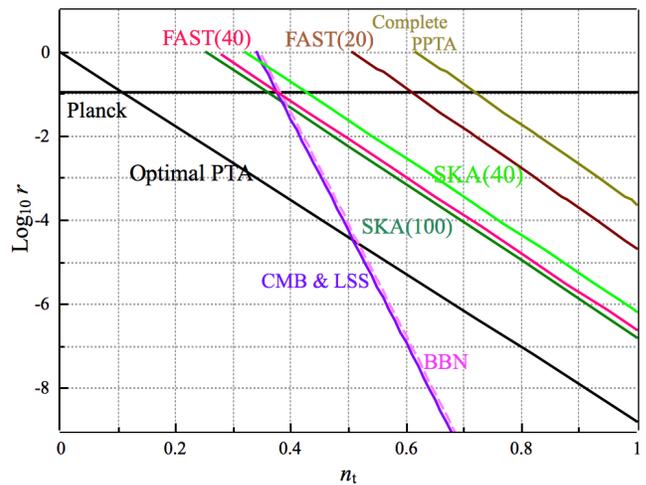}
   \end{center}
\caption{\small The constraints on $r-n_t$ space with 95\% by future observations compared with current observations from CMB and BBN. Here, the PTAs consider both cosmic expansion $T_z(k, \tau_0)$ and phase transitions $T_{\rm PT}(k, \tau_0)$. The lines of CMB and BBN are copied from Fig.\ \ref{BBN_LIGO_CMB_phase_transition}, while those of PTAs from Fig.\ \ref{impact_of_pt_on_ptas}.}
\label{all_constraints_comparision}
\end{figure}

For PTAs we only deal with the effects of $T_{\rm PT}(k, \tau_0)$ due to QCD transition and ignore the damping effects of $e^{+}e^{-}$ annihilation and SUSY breaking, as $e^{+}e^{-}$ annihilation affects RGWs in the range of $f<1.7\times10^{-12}$ Hz (or, equivalently, $0.1$ MeV) and SUSY phase transition in the range of $f>3.1\times10^{-5}$ Hz (or $10^6$ MeV). We add the transfer function $T_{\rm PT}(k, \tau_0)$ and $T_\nu(k, \tau_0)$ to the total transfer function, and follow the procedures described in Section \ref{future_pta_projects} to obtain SNR and the constraints on $r$ and $n_t$ respectively.

In Fig.\ \ref{snr1_under_phase_transition}, the red and orange lines show the SNR for RGWs further damped by $T_{\rm PT}(k, \tau_0)$ in FAST plans. Obviously, compared with the cases which only contain $T_z(k, \tau_0)$, SNR decreases for both plans when additional cosmic phase transitions are considered. The decline of SNR is consistent with our intuition, as additional damping could surely lead to weaker RGW signals and smaller SNR. In each observation, when $n_t>0.6$, for $r=0.1$, SNR decreases by an almost constant amount: FAST(40) decreases one unit, while FAST(20) by amount of 0.6. So cosmic phase transitions are more important in FAST(40) than in FAST(20). As $n_t$ decreases, the difference in SNR caused by $T_{\rm PT}(k, \tau_0)$ also decreases and tends to zero at $n_t=0$. Similar features appear in SKA observations too (see Fig.\ \ref{snr2_under_phase_transition}).

Fig.\ \ref{impact_of_pt_on_ptas} shows the effects of cosmic phase transitions $T_{\rm PT}(k, \tau_0)$ on parameters $r$ and $n_t$. Obviously, $T_{\rm PT}(k, \tau_0)$ raises the upper limits of all PTAs nearly by a constant factor of $2.2$ in the wide range of $n_t$. It is not difficult to understand the reason, as the big jump cased by QCD transition is the main feature in the narrow sensitive band of PTA  (see Fig.\ \ref{spectrum_of_energy_density_all_effects}).

In Fig.\ \ref{all_constraints_comparision}, we put the constraints from future PTAs and those from current observations together to get a better view. The bounds from PTAs have already included damping effects from cosmic expansion $T_z(k, \tau_0)$ and phase transitions $T_{\rm PT}(k, \tau_0)$. We see that around $n_t=0.38$, SKA(100) could do much better in limiting $r$ and become very competitive with current CMB observations from Planck and the joint analysis from CMB and LSS. In $n_t\in(0.36, 0.38)$, SKA(100) is a bit more (about a factor of $1.3$) stringent than BBN and CMB. In comparison with constraints from CMB and BBN, the role of cosmic phase transitions becomes particularly important for SKA(100), because if without $T_{\rm PT}(k, \tau_0)$ the advantage of SKA(100) could increase to about $3$ times better than CMB and BBN over a wider range. We can also find that optimal PTA could replace the dominant role of CMB and BBN in the range $n_t\in(0.11, 0.5)$ and restrict $n_t<0.23$ for $r=0.01$ and $n_t<0.34$ for $r=0.001$, which are very tight limits for a wide variety of inflation models, including slow-rolling inflation. We list the upper limits on $n_t$ in Table \ref{limits_on_nt_all_comparision} for clarity. When we fix $n_t$, we see that at $n_t=0$, CMB is still most stringent constraints even for the optimal PTA, but for a small positive tensorial index. e.g. $n_t=0.2$, the optimal PTA can require $r<0.02$. Therefore, the future constraints are much better than the present ones if the spectrum of RGWs is blue tilted (i.e. $n_t>0$). The bounds on $r$ for different indices $n_t$ are listed in Table \ref{limits_on_r_all_comparision}.

\begin{table}[!htb]
 \centering
  \begin{tabular}{c @{\extracolsep{4em}} c c}
  \hline\hline
  \multirow{2}*{$r$} & \multicolumn{2}{c}{upper limit on $n_t$}\\
  \Xcline{2-3}{0.4pt}
    &without $T_{\rm PT}(k, \tau_0)$  &  with $T_{\rm PT}(k, \tau_0)$\\
    \hline
   $\gape 0.1$  & $0.08$ &  $\gape 0.12$\\
   $0.01$  &  $0.18$  & $0.23$\\
   $0.001$  &  $0.30$  & $0.34$\\
  \hline
  \end{tabular}
 \caption{\small The best constraints on tensorial index $n_t$ with 95\% confidence level for three typical tensor-to-scalar ratios: $r=0.1, 0.01$ and $0.001$. Here we choose a normal EoS with $w=1/3$. As the optimal PTA would be the best constraint, these bounds with cosmic phase transitions $T_{\rm PT}(k, \tau_0)$ are all from optimal PTA in Fig.\ \ref{all_constraints_comparision}, while those without phase transition are read out from Fig.\ \ref{impact_of_pt_on_ptas}.}
  \label{limits_on_nt_all_comparision}
  \end{table}

\begin{table}[!htb]
 \centering
 \begin{tabular}{c @{\extracolsep{2em}} c c c c }
 \hline\hline
  $\gape n_t$ & $\gape r_{\rm cmb\&lss}$ & $r_{\rm ska(100)}$  & $\gape r_{\rm optimal\ pta}$\\
  \hline
  $0$  & $>1$  & $>1$  & $1$ \\
  $0.2$ &$>1$ & $>1$ & $0.02$ \\
  $0.4$ & $0.03$ & $0.04$ & $3.2\times10^{-4}$ \\
  $0.6$ & $1.0\times10^{-7}$  & $6.3\times10^{-4}$ & $6.0\times10^{-6}$ \\
  $0.8$ & $5.0\times10^{-13}$ & $1.0\times10^{-5}$ & $1.0\times10^{-7}$\\
  \hline
  \end{tabular}
  \caption{\small The constrains on the upper limit of $r$ with 95\% confidence for fixed $n_t$ from CMB \& LSS, SKA(100) and the optimal PTA. Note that the upper limit given by pure CMB (Planck) observation is $r<0.11$. All the constraints have considered the damping effects from cosmic expansion $T_z(k, \tau_0)$, cosmic phase transitions $T_{\rm PT}(k, \tau_0)$ and the free-streaming neutrinos $T_\nu(k, \tau_0)$. We use a normal EoS with $w=1/3$ here.}
  \label{limits_on_r_all_comparision}
  \end{table}

\subsection{Effects of relativistic free-streaming dark fluids}

The possible excess of effective number of neutrino species $N_{\rm eff}$ leads to tension between observations and theories. As we mentioned in Sec.\ \ref{introduction_of_dark_fluid}, this problem may be reconciled by assuming extra relativistic light particles, such as massless bosons proposed by \cite{Weinberg_2013} or a class of massless fermions i.e. the dark fluid in this paper. If verified, the existence of extra species of particles will bring new physics both to particle physics and cosmology.

For RGWs, we have found that dark fluid plays dual roles in the evolution of the gravitational waves (see Fig.\ \ref{spectrum_of_energy_density_all_effects}): It damps the spectrum by adding the fraction of free-streaming particles (see Fig.\ \ref{fraction_neutrinos_and_dark_fluid}). Meanwhile, it raises the RGW spectrum a bit through cosmic phase transitions (see Eq. (\ref{dark_fluid_raise_spectrum})).

\begin{figure}[!htb]
 \begin{center}
  \includegraphics[width=8.5cm]{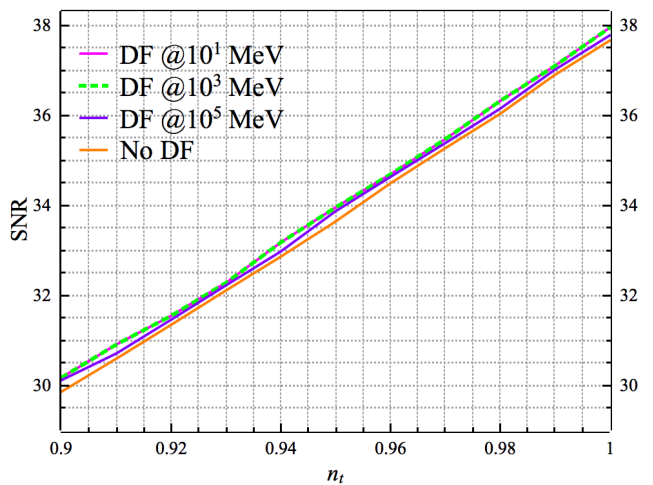}
   \end{center}
   \caption{\small The impact on SNR of SKA(100) caused by dark fluids decoupling at $10^1, 10^3$ and $10^5$ MeV compared with the case without dark fluid. Here we choose $r=0.1$ and pick out the part with obvious difference among these three cases.}
\label{dark_fluid_snr_ska}
\end{figure}

In this subsection, we will focus on the impact of dark fluid on SNR and constraints on $r-n_t$ space. As before, we also assume an EoS with $w=1/3$ before BBN. Since dark fluid could affect the spectrum through both cosmic phase transition and free-streaming gases, we should combine the transfer function $T_{\rm PT}(k, \tau_0)$, $T_{\nu}(k, \tau_0)$ and $T_z(k, \tau_0)$ to get the total transfer function in the spectrum (\ref{expanded_spectrum_of_energy_density}), then perform the similar procedures depicted in Sec.\ \ref{Detection of RGWs in the accelerating Universe} to calculate SNR of RGWs for different potential observations.

Fig.\ \ref{dark_fluid_snr_ska} shows the difference on SNR of SKA(100) caused by dark fluids decoupling at $10^1,~10^3$ and $10^5$ MeV. It is not strange to find that dark fluid can raise the SNR curve a bit ($\le0.3$), as dark fluid could raise the spectrum through phase transition and counteract the damping effect caused by its free-streaming (see Eq. (\ref{dark_fluid_raise_spectrum})). We could see this clearly in Fig.\ \ref{spectrum_of_energy_density_all_effects}: For the dark fluid decoupling at $10$ MeV (the black solid line, which is partially covered by the green solid line for $10^2$ MeV), the spectrum in the PTA band is clearly raised a bit, while for that decouples at $10^3$ MeV (the blue dashed line), the spectrum is not only raised a bit at the high frequency end (i.e. $f\sim10^{-7}$ Hz), but also damped a bit at the frequencies below QCD transition (i.e. $f\sim10^{-9}$ Hz).  While at the middle frequencies ($f\sim10^{-8}$ Hz), the two effects nearly cancel each other. For dark fluids that decouple at $10^5$ MeV, the effect of free-streaming could extend to higher frequencies, making the spectrum nearly unchanged. Therefore the SNR for the case of $10^1$ and $10^3$ MeV are raised more than that for $10^5$ MeV (the spectrum for $10^3$ MeV increases because the effect near $10^{-7}$ Hz become dominant as $n_t>0$). This is consistent with Fig.\ \ref{dark_fluid_snr_ska}, where the SNR curve for dark fluid decoupling at $10^5$ MeV (the purple solid line) is the middle one between the curve for  $10^3$ MeV (the green dotted line) and that without dark fluid (the orange solid line).

As $n_t$ goes up to $1$, the spectrum could be raised more by dark fluids, because the contribution of high frequencies become dominant for $n_t>0$. We find the largest influence on SNR caused by dark fluid for $n_t>0.9$, but the SNR of SKA(100) is only raised by less than $0.5$ for $r=0.1$. Obviously, the impact on FAST would be even smaller (similar to the cases in Fig.\ \ref{snr1_under_phase_transition}). So the influence of dark fluid could reasonably be ignored by PTAs and we cannot expect any significant changes caused by dark fluid in the $r-n_t$ space.

The reason why dark fluid plays small influence on the spectrum is simple. Current constraints on the number of species $N_{\rm eff}$ is so stringent that only one or two more kinds of extra particles are allowed, thus the contribution of dark fluid to the total radiation around the PTA band ($10^{-9}\sim10^{-7}$Hz, or equivalently, $52\sim3.9\times10^3$ MeV) is quite small, so both phase transition and free-streaming of dark fluid become insignificant.

\subsection{Effects of different equation of state $w$}
BBN could trace physics up to about 10 MeV in the history of Universe. However, the physics above this energy scale is not quite clear, for example, EoS in this range is not determined. Reheating after inflation and the appearance of massive particles may lead to quite interesting EoS, which could make the Universe evolve differently. For RGWs, these unusual EoS may greatly amplify or depress the amplitudes, making the detection hopeful or despaired (see Fig.\ \ref{spectrum_energy_density_EoS}). With the advancement in the detection of RGWs, more stringent constraints could be set to the EoS in this early stage of the Universe and limit the possible reheating physics.

Since the transfer function of unusual EoS alters the RGW spectrum dramatically (up to several orders of magnitude), we will neglect the damping effects induced by cosmic phase transitions and free-streaming particles in this subsection, and only consider effects caused by cosmic expansion $T_z(k, \tau_0)$ and unusual EoS $T_{\rm EoS}(k, \tau_0)$. Similar to the previous works \cite{Zhao_2011,LIGO_correlation}, we will consider five different EoS with $w=\infty, ~1, ~0.6, ~1/3$ and $0$ to cover the stiff, soft and radiation-like states.

\begin{figure}[!htb]
 \begin{center}
  \includegraphics[width=8.5cm]{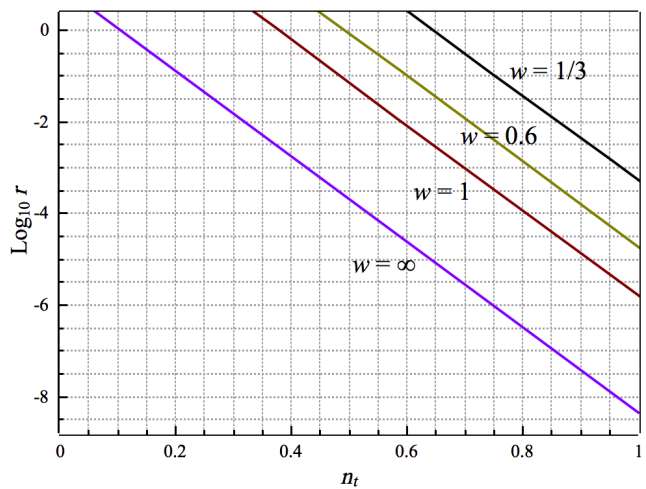}
   \end{center}
 \caption{\small The constraints on $r$ and $n_t$ with 95\% confidence level by use of NANOGrav data \cite{NANO_2015}. Here the transfer function of cosmic expansion $T_z(k, \tau_0)$ and that of unusual EoS $T_{\rm EoS}(k, \tau_0)$ are considered. Note that the constraints of EoS with $w=0$ are quite loose and exceed the scale of this plot, while the line for $w=\infty$ is linearly extended to $n_t =1$.}
 \label{constraints_on_r_nt_nano_EoS}
 \end{figure}

Firstly, we consider, when unusual EoS $T_{\rm EoS}(k, \tau_0)$ appears, to what extent the current NANOGrav data can limit the $r-n_t$ space. With procedures similar to (\ref{expresson_of_r_redshift}) and (\ref{scalar_to_tensor_ration_redshift_pt}), we could find the expression of $r$. Then by use of the NANOGrav data (see Fig.\ \ref{constraints_on_a}), we can calculate the upper limits on $r$ for the unusual EoS mentioned above. Fig.\ \ref{constraints_on_r_nt_nano_EoS} shows the constraints on $r-n_t$ space when both $T_z(k, \tau_0)$ and $T_{\rm EoS}(k, \tau_0)$ are considered. We can see that stiff EoS ($w>1/3$) could greatly suppress the possible parameter space. For example when $r=0.1$, the upper limit on $n_t$ is $0.6$ for $w=0.6$ and $0.49$ for $w=1$, both of which are more compact than that of $0.76$ for $w=1/3$. The upper limit could even be pushed to $0.21$ for $w=\infty$.

Now we consider the potential constraints from the future PTAs. Inserting the transfer function $T_{\rm EoS}(k, \tau_0)$ of unusual EoS and $T_z(k, \tau_0)$ of cosmic expansion to the total transfer function in the RGW spectrum, we could follow the steps given in Sec.\ \ref{Detection of RGWs in the accelerating Universe} to calculate the SNR and the constraint on $r-n_t$ for different observations.

\begin{figure}[!htb]
\begin{center}
 \includegraphics[width=8.5cm]{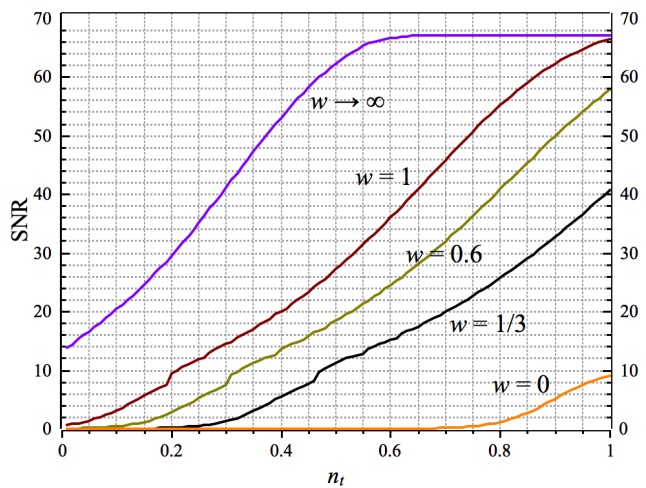}
  \end{center}
  \caption{\small The SNR of SKA(100) when different EoS before BBN are considered. Here we choose $r=0.1$ and plot SNR for five different EoS. The damping effect of cosmic expansion $T_z(k, \tau_0)$ is also included in the calculations.}
  \label{snr_ska_EoS}
  \end{figure}

Fig.\ \ref{snr_ska_EoS} shows the SNR for different EoS in the SKA(100) project. We can see that, compared with the normal EoS of radiation ($w=1/3$), stiff EoS ($w>1/3$) could greatly increase the SNR of PTAs in the whole range of $n_t\in(0.15, 1)$. SNR even starts to saturate with SNR = 67.1 at $n_t=0.6$ for the infinity stiff EoS, and reaches maximum at $n_t=1$ for $w=1$. Around $n_t=0$, SNR for $w=1$ and infinity can reach a well level. For example, when $n_t=0.07$, we have SNR=2 for $w=1$. Therefore, SKA(100) is still very sensitive to the EoS above $w>1$. If the EoS before BBN is dominated by matter of $w=0$, then the SNR will be dramatically damped in the whole range of $n_t\in(0, 1)$ and the maximum of SNR is $8.9$ at $n_t=1$. So, in conclusion, RGWs are quite sensitive to the EoS before BBN and could provide good probes into the interesting physics in this epoch.

\begin{figure}[!htb]
 \begin{center}
  \includegraphics[width=8.5cm]{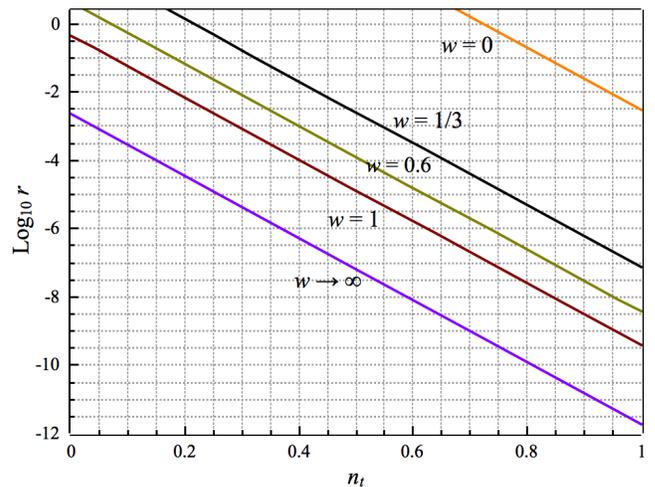}
   \end{center}
  \caption{\small Constraints on $r-n_t$ space with 95\% confidence level in SKA(100)  for different EoS ($w=\infty, 1, 0.6, 1/3$ and $0$) before BBN epoch. Here we consider damping effects from cosmic expansion $T_z(k, \tau_0)$ and that from unusual EoS $T_{\rm EoS}(k, \tau_0)$ before BBN. }
   \label{r_nt_FAST_EoS}
   \end{figure}

Fig.\ \ref{r_nt_FAST_EoS} shows the constraints on $r-n_t$ for SKA(100) when different EoSs are considered. We could see that, stiff EoS with $w>1/3$ makes the constraints more stringent than the case of $w=1/3$, while the matter-like EoS ($w=0$) damps the spectrum so much that the constraint becomes very loose. Explicitly, at $r=0.1$, the upper limit on $n_t$ given by $w=\infty$ is less than $0$, while $w=1$ gives $0.08$, and $w=0.6$ gives $0.18$, all of which are more stringent than $0.32$ given by normal radiation-like EoS ($w=1/3$), but if $w=0$, the upper limit on $n_t$ is near 1. So we can conclude that stiff EoS could facilitate the constraints of RGW parameters. For the case of $r=0.1$, $0.01$ and $0.001$, we list the constraints on $n_t$ in Table \ref{nt_fixed_r_EoS}. \\

\begin{table}[!htb]
 \centering
 \begin{tabular}{c @{\extracolsep{4em}} c c c }
 \hline\hline
  \multirow{2}*{$w$} & \multicolumn{3}{c}{upper limit of $n_t$}\\
     \Xcline{2-4}{0.4pt}
   & $r=0.1$  & $r=0.01$  &  $r=0.001$\\
\hline
$\infty$ & $<0$ & $<0$ & $0.04$\\
$1$  &  $0.08$  & $0.18$  & $0.29$\\
$0.6$ & $0.18$  &  $0.29$  &  $0.40$\\
$1/3$  &  $0.32$  &  $0.44$  &  $0.55$\\
$0$   & $0.84$ & $0.94$  & $>1$\\
\hline
\end{tabular}
\caption{\small Upper limit of $n_t$ for different EoS given by SKA(100) with 95\% confidence level when $r$ is chosen to be the three typical values: $r=0.1, 0.01$ and 0.001. The damping effect from cosmic expansion $T_z(k, \tau_0)$ is also included here.}
\label{nt_fixed_r_EoS}
\end{table}

\section{Conclusions\label{section4}}

A stochastic background of relic gravitational waves, generated during the early inflationary stage, is a necessity dictated by general relativity and quantum mechanics. The spectrum of RGWs directly depends on the inflationary physics. So, it is always treated as the smoking-gun evidence of inflation. In addition, in the post-inflation stage, various cosmic phase transitions and relativistic free-steaming gases, also left imprints on the RGW spectrum. For this reason, RGWs also provide the cleanest way to probe the physics in the post-inflation epoch.

Detections of RGWs have been carried on by different experimental methods. In the median frequency range $f \in (10^{-9},~10^{-7})$ Hz, the detection is by analyzing the timing residual of the millisecond pulsars. Recently, all three PTA groups (PPTA, EPTA and NANOGrav) reported the latest constraints on the GW background. In this paper, by considering the current and the potential future PTA observations, we investigated the constraints on the RGWs and the inflationary parameters in the general cosmological scenario. In particular, we studied the effects of cosmic phase transitions and various relativistic free-streaming fluids, which had been neglected in all the previous works.

Cosmic phase transitions, including $e^{+}e^{-}$ annihilation, QCD transition and SUSY breaking, damp the RGW spectrum in the frequency range $f>10^{-10}$Hz, which is exactly the sensitivity range of PTA method. Taking into account the damping effects caused by all physical transitions, we find the upper limit of the tensor-to-scalar ratio $r$ increases by a factor $\sim 2$ for any given spectral index $n_t$. In the standard cosmological scenario, for current NANOGrav constraints with $n_t=0$, the upper limit of $r$ increases from $8.5\times 10^{5}$ to $2.0\times10^{6}$. While, for the future SKA(100), if $r=0.1$, we find the detection of RGWs is possible only if $n_t>0.36$, instead of $n_t>0.32$.

The relativistic free-streaming gases, including neutrinos and some unknown dark fluids in the Universe, influenced the evolution of RGWs in the radiation-dominant stage, which significantly damped the RGWs spectrum at the frequency range $f>10^{-16}$Hz. By analysis, we find this effect is very small in frequency $f \in (10^{-9},~10^{-7})$ Hz for both neutrinos and dark fluids. Even for the future SKA(100) project, this effect seems impossible to be detected.

In the general cosmological scenario, the effective EoS $w$ of the cosmic fluid can be different from $1/3$ in the pre-BBN stage. We find that, the value of $w$ greatly affects the RGWs spectrum, as well as their detection by pulsar timing arrays. For the future SKA(100) project, if $r=0.01$ is set, we find the detection is possible if $n_t>0.44$ for the standard model with $w=1/3$. However, if $w=1$, they can be detected only if $n_t>0.18$. So, the stiff EoS could significantly decreases the difficulties of RGW detection.

At the end of this paper, we should mention that in addition to the detection methods above, RGWs have also been constrained by some other observational and experimental efforts. In the frequency range $f\in(10^{-16},~10^{-9})$Hz, GW background produces a pattern of apparent proper motion of quasars. So, by observing the motion of quasars in the Universe, an interesting constraint $\Omega_{\rm gw}\lesssim 0.2$ in this frequency band is given in \cite{qusar}. For the RGWs with high frequency $f>10^5$ Hz, by analyzing the implications of graviton to photon conversion in the presence of large-scale magnetic fields, an upper limit $\Omega_{\rm gw}\lesssim 1$ is derived in \cite{baskaran}. Meanwhile, the experimental detections of the RGWs by various GW detectors (e.g. the cryogenic resonant bar detectors \cite{bar}, the cavity detectors MAGO\cite{mago}, the waveguide detectors \cite{waveguide}, the Gaussian maser beam detectors \cite{maser}) have also been well studied in the recent literatures.

~


{\bf Acknowledgements:} This work is supported by Project 973 under Grant No. 2012CB821804 and 2014CB845806, by NSFC No. 11173021, 11322324, 11421303, 11275187, 11421303, 11373014 and 11073005, SRFDP, CAS, the Strategic Priority Research Program ¡±The Emergence of Cosmological Structures ¡± of the Chinese
Academy of Sciences, Grant No. XDB09000000, and the
Fundamental Research Funds for the Central Universities and Scientific Research Foundation
of Beijing Normal University.


\end{document}